\documentclass[aps,prb,twocolumn,superscriptaddress,floatfix,showpacs,groupedaddress]{revtex4-1}
\usepackage{graphicx}
\usepackage{amsfonts}
\usepackage{amssymb}
\usepackage{amsmath}
\usepackage{mathrsfs}
\usepackage{txfonts}
\usepackage{lipsum}
\usepackage{color}

\DeclareMathOperator{\Tr}{Tr}
\DeclareMathOperator{\Det}{Det}
\DeclareMathOperator{\Pfaf}{Pf}
\DeclareMathOperator{\sgn}{sgn}

\begin{document}
\title{Topological spin-singlet superconductors with underlying sublattice structure}
\author{C. Dutreix}
\affiliation{Univ Lyon, Ens de Lyon, Univ Claude Bernard, CNRS, Laboratoire de Physique, F-69342 Lyon, France
}


\begin{abstract}
Majorana boundary quasiparticles may naturally emerge in a spin-singlet superconductor with Rashba spin-orbit interactions, when a Zeeman magnetic field breaks time-reversal symmetry. Their existence and robustness against adiabatic changes is deeply related, via a bulk-edge correspondence, to topological properties of the band structure.
The present paper shows that the spin-orbit may be responsible for topological transitions when the superconducting system has an underlying sublattice structure, as it appears in a dimerized Peierls chain, graphene, and phosphorene. These systems, which belong to the Bogoliubov--de Gennes class D, are found to have an extra symmetry that plays the role of the parity. It enables the characterization of the topology of the particle-hole symmetric band structure in terms of band inversions. The topological phase diagrams this leads to are then obtained analytically and exactly. They reveal that, because of the underlying sublattice structure, the existence of topological superconducting phases requires a minimum doping fixed by the strength of the Rashba spin-orbit. Majorana boundary quasiparticles are finally predicted to emerge when the Fermi level lies in the vicinity of the bottom (top) of the conduction (valence) band in semiconductors such as the dimerized Peierls chain and phosphorene. In a two-dimensional topological superconductor based on (stretched) graphene, which is semimetallic, Majorana quasiparticles cannot emerge at zero and low doping, that is, when the Fermi level is close to the Dirac points. Nevertheless, they are likely to appear in the vicinity of the van Hove singularities.
\end{abstract}

\maketitle

\section*{Introduction}

Although Dirac introduced his Lorentz invariant equation to describe relativistic fermions in a 1928 seminal work entitled \textit{The Quantum Theory of the Electron}, it also turned out to be a remarkable prediction of antimatter, as successfully confirmed a few years latter with the discovery of the positron by Anderson [\onlinecite{dirac1928quantum},\,\onlinecite{anderson1933positive}]. Thus, when a particle is ruled by the Dirac equation of motion, there exists a conjugated solution with the same mass but opposite charge: the antiparticle. Italian physicist Majorana subsequently realized that this equation allows solutions that are their own charge conjugates [\onlinecite{Majorana:1937hc}]. The neutral elementary particles they describe are their own antiparticles, which defines what is now referred to as Majorana fermions. Investigations into low-energy Majorana quasiparticles have more recently been undertaken in condensed matter physics too [\onlinecite{kitaev2001unpaired}], especially in the context of spinless $p_{x}+ip_{y}$ superconductivity, as there may be in Sr$_{2}$RuO$_{4}$, and where they may appear as entangled anyons, whose non-Abelian braiding is a promising mechanism for fault-tolerant quantum computers [\onlinecite{read2000paired,ivanov2001non,mackenzie2003superconductivity,nayak2008non}]. Even though Majorana quasiparticles were also discussed in connection to noncentrosymmetric superconductors with a mixture of singlet and triplet pairings [\onlinecite{Sato:2006kl,Sato:2009fe,Sato:2011qo}], a decisive step forward was made with pioneering proposals that only involved conventional spin-singlet superconductivity, when it is induced by proximity effect in materials with spin-orbit interactions [\onlinecite{Fu:2008qy,sau2010generic,Alicea:2010ec,klinovaja2012helical,klinovaja2013spintronics,klinovaja2013giant}]. This was followed by predictions in  one-dimensional (1D) semiconductors under time-reversal symmetry breaking Zeeman magnetic field [\onlinecite{Oreg:2010fu},\,\onlinecite{Lutchyn:2010bv}], before being confirmed in nanowires of InSb and InAs with the observations of zero-bias peaks and exponentially localized zero-energy states by Coulomb blockade spectroscopy [\onlinecite{mourik2012signatures,Das:2012aa,albrecht2016exponential}]. It has subsequently been realized that both the Zeeman field and Rashba spin-orbit could be simulated by magnetic adatoms of Fe deposited on the surface of a Pb superconductor [\onlinecite{choy2011majorana},\,\onlinecite{nadj2014observation}], which was then extended to other materials [\onlinecite{sedlmayr2015majoranas},\,\onlinecite{sedlmayr2015flat}]. In the spin-singlet superconducting materials without time-reversal symmetry mentioned above, the Majorana quasiparticles arise as zero-energy boundary modes and result from topological properties of a particle-hole symmetric Bloch Hamiltonian. In these systems, the topological properties and, \textit{a fortiori}, the existence of the Majorana quasiparticles do not depend on the strength of the Rashba spin-orbit, whose role essentially consists in protecting the superconducting gap, whereas the Zeeman magnetic field tends to close it. This is in strong analogy with the role of the intrinsic spin-orbit interactions which ensures the existence of a non-zero bulk energy gap in the quantum spin Hall phase [\onlinecite{Kane:2005fu,Kane:2005dz,Fu:2007oz}].

Here we will see that, contrary to the works discussed above, the Rashba spin-orbit may actually be responsible for topological phase transitions in spin-singlet superconductors that have an underlying sublattice structure. In order to explain why and to what extent the strength of the Rashba spin orbit influences the existence of Majorana boundary quasiparticles in multiatomic-pattern crystals, the present paper is organized as follows. Section\,\ref{Bogoliubov-de Gennes class D} provides a general prescription that allows us to apprehend the topology of the 1D and 2D Bloch band structures we will subsequently concerned with, and that belong to the Bogoliubov--de Gennes (BdG) class D. It establishes an explicit relation between the topological invariants and the band inversions that occur at some symmetric momenta of the Brillouin zone (BZ). It crucially turns out that the Rashba spin-orbit has no reason to vanish at these peculiar momenta in the case of crystals with underlying sublattice structure. This suggests that this spin-flip process may have a direct influence over the topological phase transitions. This is the purpose of Section\,\ref{Application to diatomic-pattern crystals}, which also focuses on some specific applications in 1D and 2D multiatomic-pattern crystals such as the dimerized Peierls chain, (stretched) graphene, and phosphorene. It explicitly emphasizes the effects of the Rashba spin-orbit strength through topological phase diagrams. They reveal that Majorana boundary quasiparticles are likely to emerge at the bottom (top) of the conduction (valence) band in 1D and 2D semiconductors such as the dimerized Peierls chain and phosphorene. They also demonstrate that the spin-orbit requires the Fermi level to be fixed away from the Dirac points in a 2D semimetal such as graphene, and that chiral Majorana modes are allowed to emerge in the vicinity of the van Hove singularities.

\section{Bogoliubov-de Gennes class D}
\label{Bogoliubov-de Gennes class D}

\subsection{$\Pi$ symmetry}
Noninteracting electrons in a crystal with discrete translation symmetry can be described in terms of Bloch band structures, which is represented by a $M \times M$ Hamiltonian matrix $\mathcal{H}({\bf k})$. The dimension of wave vector ${\bf k}$ is arbitrary if not specified. Here we aim to discuss the Bloch band structures that belong to the BdG class D in the Altland and Zirnbauer symmetry table [\onlinecite{Altland:1997fy}]. Consequently, time-reversal and chiral symmetries are assumed to be broken. Nonetheless, the Bloch band structure still has particle-hole symmetry (PHS) and the associated charge-conjugation operator squares to plus identity. The band structure is additionally assumed to have an extra symmetry which, for some reasons that will become clearer shortly, is referred to as $\Pi$ symmetry ($\Pi$S) with reference to parity (or inversion) symmetry. These two symmetries are defined as follows:
\begin{align}
\label{PHS}
\mathcal{C}\,\mathcal{H}({\bf k})\,\mathcal{C}^{-1} &= -\mathcal{H}^{*}(-{\bf k}) ~~~~~~\text{with}~~~~~~ \mathcal{C}^{2} = +1 \,, \\
\label{PiS}
\mathcal{P}\,\mathcal{H}({\bf k})\,\mathcal{P}^{-1} &= \,\,\,\,\mathcal{H}(-{\bf k}) ~~~~~~\,\,\text{with}~~~~~~ \mathcal{P}^{2} = +1 \,,
\end{align}
where both $\mathcal{C}$ and $\mathcal{P}$ are unitary operators that anticommute with each other
\begin{align}
\label{Anticommutation}
\{\mathcal{P},\mathcal{C}\}=0 \,.
\end{align}

As a consequence of PHS (\ref{PHS}), the eigenstates of $\mathcal{H}({\bf k})$ come in pairs at opposite momenta with opposite energies:
\begin{align}
\mathcal{H}({\bf k}) \, |u_{m}({\bf k})\rangle &= \mathcal{E}_{m}({\bf k}) \, |u_{m}({\bf k})\rangle \,, \notag \\
\mathcal{H}(-{\bf k}) \, \mathcal{C} |u_{m}^{*}({\bf k})\rangle &= - \mathcal{E}_{m}({\bf k}) \, \mathcal{C} |u_{m}^{*}({\bf k})\rangle \,,
\end{align}
where $| u_{m}({\bf k}) \rangle$ is the orbital part of the $m$-th Bloch state. Besides, it is implied that they form a complete orthonormal basis of the Hilbert space
\begin{align}\label{completeness relation}
\sum_{m=1}^{M}  |u_{m}({\bf k})\rangle \langle u_{m}({\bf k})| = 1 \,.
\end{align}
The Bloch spectrum is particle-hole symmetric and $M$ is necessarily even. The zero of energies in the BdG quasiparticle spectrum is defined with respect to the chemical potential, as usual.

$\Pi$S (\ref{PiS}) implies that the eigenstates of $\mathcal{H}$ come in pairs at opposite momenta but with the same energy:
\begin{align}
\mathcal{H}({\bf k}) \, |u_{m}({\bf k})\rangle &= \mathcal{E}_{m}({\bf k}) \, |u_{m}({\bf k})\rangle \,, \notag \\
\mathcal{H}(-{\bf k}) \, \mathcal{P} |u_{m}({\bf k})\rangle &= \mathcal{E}_{m}({\bf k}) \, \mathcal{P} |u_{m}({\bf k})\rangle \,.
\end{align}
$\Pi$S also suggests the definition of special symmetry points $\mathbf{\Gamma}_{i}$ at which $\mathcal{H}$ remains invariant under operator $\mathcal{P}$. These are the momenta satisfying $\mathbf{\Gamma}_{i}=\mathbf{G}/2$, where $\mathbf{G}$ is a vector that belongs to the reciprocal Bravais lattice. This leads, along with the $\mathbf{G}$ periodicity of $\mathcal{H}({\bf k})$, to the commutation relation
\begin{align}
\label{Commutation}
[\mathcal{P},\,\mathcal{H}(\mathbf{\Gamma}_{i})]=0 \,.
\end{align}
Thus, there exists a commune basis of eigenvectors such that
\begin{align}
\mathcal{H}(\mathbf{\Gamma}_{i}) \, |u_{m}(\mathbf{\Gamma}_{i})\rangle &= \mathcal{E}_{m}(\mathbf{\Gamma}_{i}) \, |u_{m}(\mathbf{\Gamma}_{i})\rangle \,, \notag \\
\mathcal{P} \, |u_{m}(\mathbf{\Gamma}_{i})\rangle &= \pi_{m}(\mathbf{\Gamma}_{i}) \, |u_{m}(\mathbf{\Gamma}_{i})\rangle \,,
\end{align}
while the anticommutation relation (\ref{Anticommutation}) additionally implies
\begin{align}
\mathcal{H}(\mathbf{\Gamma}_{i}) \, \mathcal{C} |u_{m}^{*}(\mathbf{\Gamma}_{i})\rangle &= - \mathcal{E}_{m}(\mathbf{\Gamma}_{i}) \, \mathcal{C} |u_{m}^{*}(\mathbf{\Gamma}_{i})\rangle \, \notag \\
\mathcal{P} \, \mathcal{C} |u_{m}^{*}(\mathbf{\Gamma}_{i})\rangle &= -\pi_{m}(\mathbf{\Gamma}_{i}) \, \mathcal{C} |u_{m}^{*}(\mathbf{\Gamma}_{i})\rangle \,.
\end{align}
As a result, the eigenstates come in pairs with opposite energies and opposite parities $\pm\pi_{m}$ at every symmetry point $\mathbf{\Gamma}_{i}$. Since operator $\mathcal{P}$ is unitary and squares to plus the identity operator, its eigenvalues lie on the unit circle and are real, which implies $\pi_{m}(\mathbf{\Gamma}_{i})=\pm1$.

\subsection{Parity product $\delta_{i}$}
The parity $\pi_{m}({\bf\Gamma}_{i})$ can be used to label every energy band at the symmetry points ${\bf\Gamma}_{i}$. In virtue of PHS (\ref{PHS}), the knowledge of all the parities of the negative-energy bands ($\mathcal{E}_{m}<0$) is sufficient to recover the parity of the positive-energy bands ($\mathcal{E}_{m}>0$) and vice versa. This enables us to focus on the parity product of all the negative-energy bands
\begin{align}
\label{Parity Product 1}
\delta_{i}=\prod_{E_{m}<0}\pi_{m}({\bf\Gamma}_{i}) \,.
\end{align}
Initially introduced by Fu and Kane in connection to topological insulators with inversion symmetry [\onlinecite{Fu:2007oz}], this quantity has subsequently been generalized by Sato to odd-parity superconductors [\onlinecite{Sato:2010lh}], and also been discussed in the context of Floquet topological insulators [\onlinecite{dutreix2016laser}]. Of course, an equivalent definition holds for the positive-energy bands too. The parity product cannot change continuously, since it only takes integer values, namely $\delta_{i}=\pm1$. In order to change, the bulk energy-gap must close at a symmetry point $\mathbf{\Gamma}_{i}$. Like this, at least two particle-hole symmetric bands become degenerate at zero energy and can change parities, meanwhile $\delta_{i}$ becomes ill defined. Such a parity change defines a band inversion, and we will see in what follows that it may be associated to a change of the Bloch band structure topology. An alert reader may already recognize here the symmetry-protected topological feature of $\delta_{i}$, which cannot change continuously, and can only change when the particle-hole symmetric gap closes at zero energy.

\subsection{The $\Pi$-basis}
We define the $\Pi$-basis as the basis that diagonalizes operator $\mathcal{P}$ with the new representation $\tilde{\mathcal{P}}=1_{M/2} \otimes \tau_{3}$, where $1_{M/2}$ denotes the $M/2 \times M/2$ identity matrix, and $\tau_{3}$ is the third Pauli matrix that refers to the subspaces of positive and negative parities. The anticommutation relation (\ref{Anticommutation}) can explicitly be written as
\begin{align} 
\left( \begin{array}{ll} 
1_{M/2} & 0 \\
0 & -1_{M/2}
\end{array} \right)
\left( \begin{array}{ll} 
\mathcal{\tilde{C}}_{1} & \mathcal{\tilde{C}}_{2} \\
\mathcal{\tilde{C}}_{3} & \mathcal{\tilde{C}}_{4}
\end{array} \right)
\left( \begin{array}{ll} 
1_{M/2} & 0 \\
0 & -1_{M/2}
\end{array} \right)=-
\left( \begin{array}{ll} 
\mathcal{\tilde{C}}_{1} & \mathcal{\tilde{C}}_{2} \\
\mathcal{\tilde{C}}_{3} & \mathcal{\tilde{C}}_{4}
\end{array} \right) \,. \notag
\end{align}
This obviously requires the charge conjugation operator to have a block off-diagonal representation that is
\begin{align}
\mathcal{\tilde{C}}=
\left( \begin{array}{ll} 
0 & \mathcal{\tilde{C}}_{2} \\
\mathcal{\tilde{C}}_{3} & 0
\end{array} \right) \,,
\end{align}
where $\tilde{\mathcal{C}}^{2}=+1$ tells us that $\mathcal{\tilde{C}}_{2}=\mathcal{\tilde{C}}_{3}^{-1}$.

In a similar way, the commutation relation (\ref{Commutation}) requires the Hamiltonian matrix to have a block diagonal representation in the $\Pi$ basis, namely
\begin{align}
\mathcal{\tilde{H}}(\mathbf{\Gamma}_{i})=
\left( \begin{array}{ll} 
\mathcal{\tilde{H}}_{1}(\mathbf{\Gamma}_{i}) & 0 \\
0 & \mathcal{\tilde{H}}_{4}(\mathbf{\Gamma}_{i})
\end{array} \right) \,.
\end{align}
And PHS (\ref{PHS}) is finally responsible for
\begin{align}
\mathcal{\tilde{H}}(\mathbf{\Gamma}_{i})=
\left( \begin{array}{ll} 
\mathcal{\tilde{H}}_{1}(\mathbf{\Gamma}_{i}) & ~~~~~0 \\
0 & -\left(\mathcal{\tilde{C}}_{3}\mathcal{\tilde{H}}_{1}(\mathbf{\Gamma}_{i})\mathcal{\tilde{C}}_{3}^{-1}\right)^{*}
\end{array} \right) \,.
\end{align}
Thus, the Hamiltonian matrix is block diagonal in the $\Pi$-basis and the fact that its eigenstates come in pairs with opposite energies and opposite parities becomes explicit. Indeed, the eigenstates belong to two distinct subspaces that refer to the positive and negative parities.

Besides, the parity product of the negative-energy bands as defined in Eq.\,(\ref{Parity Product 1}) turns out to be equivalent to the sign product of the energies with positive parities, meaning
\begin{align}
\delta_{i}=(-1)^{M/2}\prod_{\pi_{m}>0}\sgn \mathcal{E}_{m}({\bf\Gamma}_{i}) \,.
\end{align}
Remember that $M$ is necessarily even under PHS. In the $\Pi$ basis, the energy product of positive-parity bands is now given by a block determinant, so that the parity product can finally be rewritten as
\begin{align}\label{Parity Product 3}
\delta_{i}=(-1)^{M/2}\sgn [\Det \tilde{\mathcal{H}}_{1}({\bf\Gamma}_{i})] \,.
\end{align}
This expression turns out to be very practical, as it provides a relation between the parity product and the system parameters involved in the Bloch Hamiltonian matrix at the symmetry points ${\bf\Gamma}_{i}$. Importantly, it is not necessary to solve $M$ coupled secular equations to obtain the spectrum and eigenstates of $\mathcal{H}({\bf\Gamma}_{i})$, before evaluating their parity under operator $\mathcal{P}$ and computing parity $\delta_{i}$. Instead, it can be apprehended through the simpler calculation of a $M/2 \times M/2$ determinant.

\subsection{Sewing matrix and Berry connection}
One defines the sewing matrix $\mathcal{B}$ associated to all energy bands, i.e., those of negative and positive energies, as
\begin{align}
\mathcal{B}_{mn}({\bf k}) =  \langle u_{m}(-{\bf k}) | \mathcal{P} \mathcal{C} | u_{n}^{*} (-{\bf k}) \rangle \,.
\end{align}
At the symmetric points, charge conjugation operator $\mathcal{C}$ does not commute with $\mathcal{H}({\bf\Gamma}_{i})$ and its eigenvalues are not good quantum numbers. As we will see, the introduction of operator $\mathcal{P}$ in the sewing matrix allows us to label the energy bands with parities. The sewing matrix is unitary, i.e., $\mathcal{B}^{-1}({\bf k}) =  \mathcal{B}^{\dagger}({\bf k})$, and two particle-hole symmetric states are related to one another by
\begin{align}
\mathcal{C} | u_{m}^{*}(-{\bf k}) \rangle = \sum_{n=1}^{2N} \mathcal{B}_{mn}({\bf k}) \mathcal{P}^{\dagger} | u_{n}({\bf k}) \rangle \,.
\end{align}
As detailed in Appendix\,\ref{Appendix Antisymmetric sewing matrix}, the transpose of the sewing matrix verifies $\mathcal{B}^{T}_{mn}({\bf k})=-\mathcal{B}_{mn}({\bf k})$ or equivalently $\mathcal{B}^{\dagger}({\bf k}) = - \mathcal{B}^{*}({\bf k})$. Therefore, the sewing matrix is antisymmetric for all ${\bf k}$, and its Pfaffian can be defined. Note that the derivation above involves the property $\mathcal{C}=\mathcal{C}^{\dagger}$, which comes from $\mathcal{C}^{2}=+1$, along with the unitary condition $\mathcal{C}^{\dagger}\mathcal{C}=+1$. This is what makes the sewing matrix antisymmetric. Indeed the condition $\mathcal{C}^{2}=-1$ would lead to a symmetric sewing matrix instead.
At the symmetry points ${\bf \Gamma_{i}}$, the sewing matrix becomes block off-diagonal and can be written under the following form:
\begin{align}
\mathcal{B}( {\bf \Gamma_{i}} ) = 
\left( \begin{array}{ll} 
0 & \pi( {\bf \Gamma_{i}} ) \\
-\pi( {\bf \Gamma_{i}} ) & 0
\end{array} \right) \,,
\end{align}
where $\pi( {\bf \Gamma_{i}} )$ is a $M/2 \times M/2$ matrix whose components are given by $\pi_{mn}( {\bf \Gamma_{i}} )=\pi_{m}( {\bf \Gamma_{i}} )\,\delta_{mn}$. The Berry connection over all the $M$ energy bands, that is, $\mathcal{A}\left({\bf k}\right)=-i \sum_{m} \langle u_{n}({\bf k}) | \nabla_{\bf k} | u_{n}({\bf k}) \rangle$, can be expressed in terms of the sewing matrix as
\begin{align}
\label{Ln Determinant 2}
\mathcal{A}\left({\bf k}\right) + \mathcal{A}\left(-{\bf k}\right) &= -i \nabla_{\bf k} \ln \Det \mathcal{B}\left({\bf k}\right) \,,
\end{align}
whereas the Berry connections of the negative- and positive-energy bands are respectively related to one another in the following way: $\mathcal{A}^{-}({\bf k}) = \mathcal{A}^{+}(-{\bf k})$. This implies $\mathcal{A}\left({\bf k}\right)=\mathcal{A}\left(-{\bf k}\right)$ and relation (\ref{Ln Determinant 2}) can finally be rewritten as
\begin{align}\label{Berry Connection Relation}
\mathcal{A}({\bf k}) = -\frac{i}{2} \nabla_{\bf k} \ln \Det \mathcal{B} ({\bf k}) \,.
\end{align}
Details of the derivations above may be found in Appendix\,\ref{Appendix Berry connection}.

\subsection{$\mathbb{Z}_{2}$ topological invariant in 1d}
In one dimension, the BdG symmetry class D is characterized by a $\mathbb{Z}_{2}$ topological invariant [\onlinecite{schnyder2008classification}], namely $\exp[i\gamma_{BZ}]$ where $\gamma_{BZ}$ is known as Berry or Zak phase [\onlinecite{berry1984quantal},\,\onlinecite{zak1989berry}], i.e., a gauge-invariant geometrical phase picked up by the wavefunctions of negative-energy bands along the 1D Brillouin zone (BZ). It satisfies
\begin{align}
\label{Zak Phase}
\gamma_{BZ} &= -i \ln \left[ \Pfaf \mathcal{B}\left({\bf \Gamma_{0}}\right) \, \Pfaf \mathcal{B}\left({\bf \Gamma_{1}}\right) \right] \,,
\end{align}
when using Eq.\,(\ref{Berry Connection Relation}), as shown in Appendix\,\ref{Appendix topological invariant in 1d}. This subsequently leads to
\begin{align}
\label{Z2 Invariant}
e^{i\gamma_{BZ}} 
&= \delta_{0} \, \delta_{1} = \sgn [\Det \tilde{\mathcal{H}}_{1}({\bf\Gamma}_{0})] \, \sgn [\Det \tilde{\mathcal{H}}_{1}({\bf\Gamma}_{1})] \,.
\end{align}
Therefore, the $\mathbb{Z}_{2}$ topological invariant can be connected to the parity products defined at the symmetry points of the BZ. It is exactly known from the calculations of two $M/2 \times M/2$ determinants when the BdG band structure is $\Pi$-symmetric. Since $\delta_{0} \, \delta_{1} = \pm 1$, the Zak phase is necessarily $\pi$-quantized. In particular the relation $\delta_{0} \, \delta_{1}=-1$ requires $\gamma_{BZ}=\pi~[2\pi]$, and means that the system lies in a topological superconducting phase characterized by Majorana boundary quasiparticles at zero energy.

\subsection{$\mathbb{Z}$ topological invariant in 2d}
In two dimensions, the BdG symmetry class D is characterized by a $\mathbb{Z}$ topological invariant [\onlinecite{schnyder2008classification}], namely a first Chern number $\nu$. As detailed in Appendix\,\ref{Appendix topological invariant in 2d}, its definition involves the Berry curvature $\mathcal{F}^{-}({\bf k})=\nabla_{{\bf k}} \times \mathcal{A}^{-}({\bf k})$, which implies
\begin{align}\label{Chern number}
\nu &= \frac{1}{\pi} \int_{S} d^{2}k ~ \mathcal{F}^{-}\left({\bf k}\right) \,,
\end{align}
where $S$ refers to half the two-dimensional BZ, as illustrated in Fig\,\ref{Lattices}. It is outlined by an oriented path denoted $\mathscr{C}$. Besides, the Berry phase along that path is given by
\begin{align}
\label{Berry Phase}
\gamma_{\mathscr{C}}&= -i \ln \frac{\Pfaf \mathcal{B} \left({\bf \Gamma_{1}}\right)}{\Pfaf \mathcal{B} \left({\bf \Gamma_{0}}\right)}\frac{\Pfaf \mathcal{B} \left({\bf \Gamma_{3}}\right)}{\Pfaf \mathcal{B} \left({\bf \Gamma_{2}}\right)} ~.
\end{align}
Details are provided in Appendix\,\ref{Appendix topological invariant in 2d}. Similar derivations as the ones done for the 1D case straightforwardly lead to
\begin{align}
e^{i\gamma_{\mathscr{C}}} &= \prod_{i=0}^{3} \sgn [\Det \tilde{\mathcal{H}}_{1}({\bf\Gamma}_{i})] \,.
\end{align}
When the spectrum is not gapped, as it may be the case for spinless superconductivity, $\exp[i\gamma_{\mathscr{C}}]=-1$ implies that the Berry phase satisfies $\gamma_{\mathscr{C}}=\pi~[2\pi]$, and that there are an odd number of nodal points within the closed path $\mathscr{C}$. When the spectrum is gapped, however, Stokes theorem provides a relation between Eq.\,(\ref{Chern number}) and Eq.\,(\ref{Berry Phase}), which results in
\begin{align}
(-1)^{\nu} = \prod_{i=0}^{3} \sgn [\Det \tilde{\mathcal{H}}_{1}({\bf\Gamma}_{i})] \,.
\end{align}
Therefore, $\Pi$S does not lead to the exact value of the $\mathbb{Z}$ topological invariant. Nonetheless, $(-1)^{\nu}=-1$ tells that the Chern number is odd and necessarily non-zero, so that there exists symmetry-protected Majorana modes at the boundaries.

Because there exists a simple relation between band inversion and Bloch band structure topology, we from now on refer to band inversions that yield a topology change as topological band inversions.

\section{Application to multiatomic-pattern crystals}
\label{Application to diatomic-pattern crystals}

%
\subsection{Tight-binding Hamiltonians}
\begin{figure}[t]
\includegraphics[trim = 0mm 80mm 0mm 80mm, clip, width=8cm]{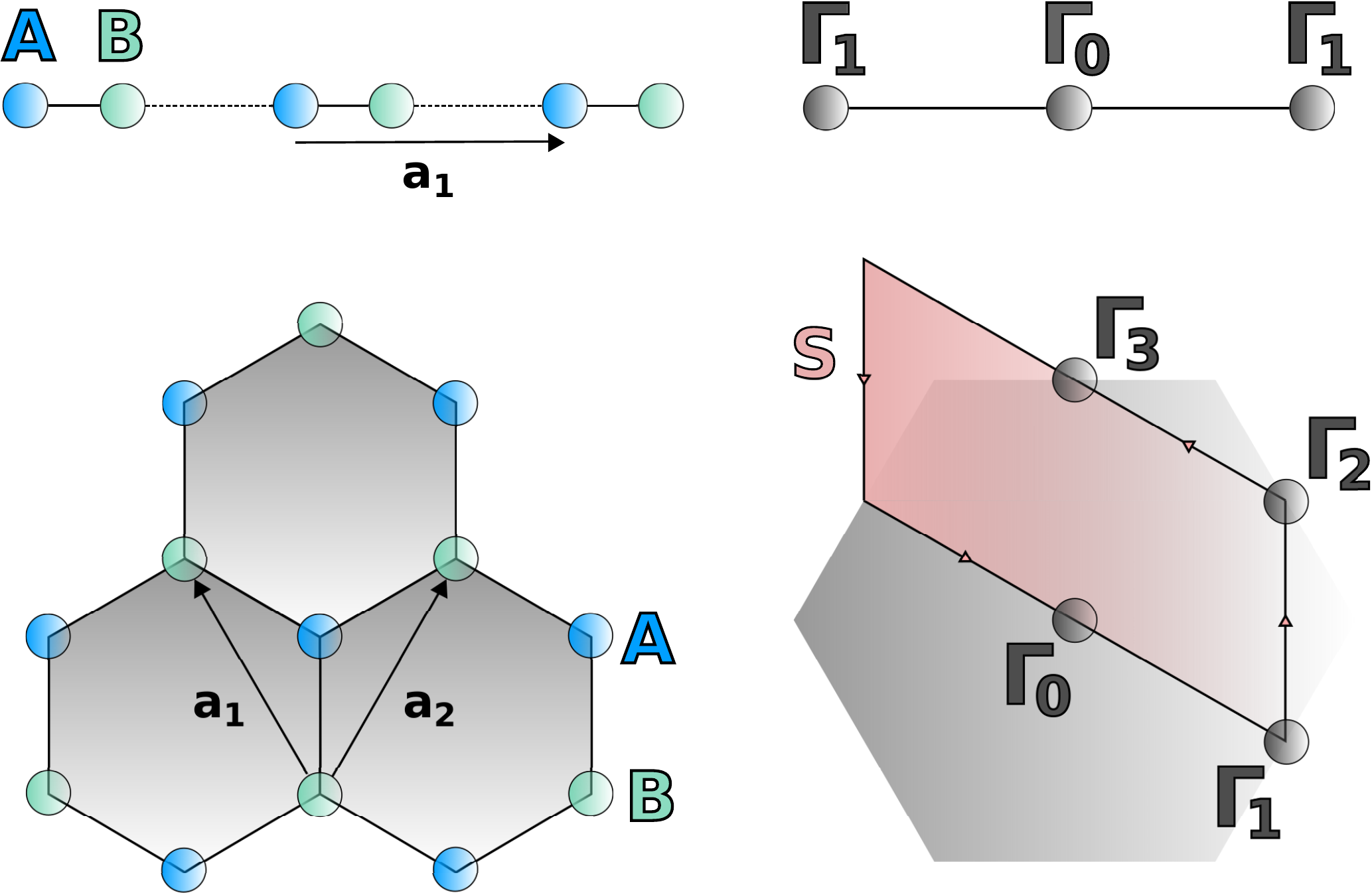}
\caption{\small (Color online) Illustration of two diatomic-pattern lattices (left) and their Brillouin zones with the symmetry points ${\bf \Gamma}_{i}$ (right). The vectors that span the Bravais lattice are denoted ${\bf a}_{i}$, while $S$ refers to the oriented surface that encloses half the 2D Brillouin zone.}
\label{Lattices}
\end{figure}

Now let us consider Bloch electrons in a crystal whose periodic structure consists of a 1D or 2D Bravais lattice with two sites per unit cell. The two nonequivalent sites define two sublattices that are referred to as sublattice A and sublattice B, as illustrated in Fig.\,\ref{Lattices}. The Bloch electrons are described within a tight-binding approach by the following Hamiltonian:
\begin{align}
\label{Kinetic Hamiltonian}
{\cal{H}}_{0}& = \sum_{{\bf k},\sigma} t\left({\bf k}\right) \left( a^{\dagger}_{{\bf k}\sigma}b_{{\bf k}\sigma} + b^{\dagger}_{{\bf k}\sigma}a_{{\bf k}\sigma} \right) \notag \\
&+\sum_{{\bf k},\sigma} \mu\left({\bf k}\right) \left( a^{\dagger}_{{\bf k}\sigma}a_{{\bf k}\sigma} + b^{\dagger}_{{\bf k}\sigma}b_{{\bf k}\sigma} \right) \,,
\end{align}
where $t$ refers to bipartite processes, namely intersublattice processes such as nearest-neighbor hopping, while $\mu$ describes the chemical potential and intrasublattice hopping processes. These are functions of the momentum ${\bf k}$, which are not specified yet. What must be specified, however, is that the sublattice structure of the crystal allows a gauge choice in the definition of the Fourier transform~[\onlinecite{bena2009remarks}], and Eq.\,(\ref{Kinetic Hamiltonian}) relies on the definition that makes the Bloch Hamiltonian periodic, i.e., $t\left({\bf k}+{\bf G}\right)=t\left({\bf k}\right)$ and $\mu\left({\bf k}+{\bf G}\right)=\mu\left({\bf k}\right)$ when ${\bf G}$ is a vector that belongs to the reciprocal Bravais lattice. The fermionic operator $a_{{\bf k}\sigma}$ ($b_{{\bf k}\sigma}$) annihilates an electron with momentum {\bf k} and spin $\sigma$ on sublattice A (B). Importantly, Hamiltonian ${\cal{H}}_{0}$ is invariant by inversion symmetry as long as no mass term of the form $m\left({\bf k}\right)\left( a^{\dagger}_{{\bf k}\sigma}a_{{\bf k}\sigma} - b^{\dagger}_{{\bf k}\sigma}b_{{\bf k}\sigma} \right)$ is considered. Such a mass term $m$ arises for example in the tight-binding descriptions of boron nitride and of the anomalous Hall effect in graphene~[\onlinecite{Haldane:1988bs}]. Therefore, ${\cal{H}}_{0}$ is a reasonable description that explains the electronic properties of 1D organic semiconductors, graphene, and phosphorene at low energy [\onlinecite{su1979solitons,katsnelson2012graphene,rudenko2014quasiparticle}]. As it will be discussed in details later on, inversion symmetry turns out to be crucial to explicitly build $\mathcal{P}$ operator as introduced in Sec.\,\ref{Bogoliubov-de Gennes class D} and then accessing the topological properties of the Bloch band structure.

As already touched on in Introduction, the quest of Majorana fermions in condensed matter physics naturally involves superconductivity, since the Bogoliubov de--Gennes quasiparticles are collective excitations of electrons and holes. Because they are their own anti-quasiparticles, Majorana quasiparticles are neutral objects and, thus, appear as zero-energy boundary modes within the particle-hole symmetric energy gap. In order to investigate the effects of the strength of the Rashba spin-orbit, we now follow the prescriptions discussed in Introduction. A Zeeman splitting potential $V_z$ is simulated as follows:
\begin{align}
{\cal{H}}_{Z}&= \sum_{{\bf k},\sigma}  \sigma\,V_z \left(a^{\dagger}_{{\bf k}\sigma}a_{{\bf k}\sigma} + b^{\dagger}_{{\bf k}\sigma}b_{{\bf k}\sigma}\right) \,.
\label{Zeeman Hamiltonian}
\end{align}
The Zeeman splitting may \textit{a priori} arise from a perpendicular magnetic field, but the latter would be responsible for orbital depairing that would reduce superconductivity in two dimensions. This detrimental issue may actually be fixed in cold atomic systems thanks to the neutrality of a s-wave superfluid [\onlinecite{sato2009non}], or by applying an in-plane magnetic field to 2D semiconductors [\onlinecite{Alicea:2010ec}]. An alternative consists in sandwiching the material between an $s$-wave superconductor and a ferromagnetic insulator. The latter, which induces a Zeeman splitting, prevents the electrons from experiencing any Lorentz force [\onlinecite{sau2010generic}].

The Rashba spin orbit arises when breaking the reflection symmetry with respect to a plane that contains the crystal. This is, for example, achieved with a perpendicular electric field or adatoms [\onlinecite{PhysRevLett.109.266801}]. The Rashba spin orbit tends to align spins in the direction defined by the nearest-neighbor vectors. This spin-flip process is characterized by
\begin{align}
{\cal{H}}_{R}&= \sum_{{\bf k},\sigma,\sigma'}  \left(\begin{array}{cc}0 & {\cal{L}}_{\uparrow\downarrow}({\bf k})\\{\cal{L}}_{\downarrow\uparrow}({\bf k}) & 0 \\\end{array}\right)_{\sigma\sigma'} a^{\dagger}_{{\bf k}\sigma}b_{{\bf k}\sigma'} +H.c.
\label{Rashba Hamiltonian}
\end{align}

Finally, spin-singlet pairing can be induced by proximity effect, but it is also likely to arise from strong electron-electron interactions in the case of doped graphene [\onlinecite{PhysRevB.75.134512,PhysRevLett.98.146801,Black-Schaffer:2012os}]. At a mean-field level, this is described by
\begin{align}\label{Superconducting Hamiltonian}
{\cal{H}}_{S}&= \sum_{{\bf k}} \Delta_0 \left(a^{\dagger}_{{\bf k}\uparrow}a^{\dagger}_{-{\bf k}\downarrow} + b^{\dagger}_{{\bf k}\uparrow} b^{\dagger}_{-{\bf k}\downarrow}\right) +H.c. \notag \\
&+ \sum_{{\bf k}}  \Delta_1({\bf k}) \left(a^{\dagger}_{{\bf k}\uparrow}b^{\dagger}_{-{\bf k}\downarrow} - a^{\dagger}_{{\bf k}\downarrow} b^{\dagger}_{-{\bf k}\uparrow}\right) +H.c.
\end{align}
The superconducting order parameters $\Delta_0$ and $\Delta_1$ denote on-site and nearest-neighbor electronic interactions, respectively. Both are considered simultaneously for more generality.

\subsection{Rashba spin-orbit at the symmetry points ${\bf \Gamma}_{i}$}
The Rashba spin-orbit is simulated here as a nearest-neighbor spin-flip hopping process that does not break time-reversal symmetry. This results in
${\cal{L}}_{\downarrow \uparrow}\left({\bf k}\right) = - {\cal{L}}_{\uparrow \downarrow}^{*}\left(-{\bf k}\right)$, 
regardless of the number of sublattices involved in the crystal. If there is a monatomic pattern with a single orbital per site, the Rashba Hamiltonian introduced in Eq\,(\ref{Rashba Hamiltonian}) reduces to
\begin{align}\label{Rashba Hamiltonian}
{\cal{H}}_{R} &= \sum_{{\bf k}}  {\cal{L}}_{\uparrow \downarrow}\left({\bf k}\right) a^{\dagger}_{{\bf k}\uparrow}a_{{\bf k}\downarrow} +
{\cal{L}}_{\downarrow \uparrow}\left({\bf k}\right) a^{\dagger}_{{\bf k}\downarrow}a_{{\bf k}\uparrow} \,.
\end{align}
Then the Hermiticity of the Hamiltonian yields the additional condition ${\cal{L}}_{\downarrow \uparrow}\left({\bf k}\right) = {\cal{L}}_{\uparrow \downarrow}^{*}\left({\bf k}\right)$ which, along with time-reversal symmetry, implies ${\cal{L}}_{\sigma \sigma'}\left({\bf k}\right) = - {\cal{L}}_{\sigma \sigma'}\left(-{\bf k}\right)$. Therefore, the Rashba spin-orbit coupling is an odd function of the momentum, 
as it also occurs due to the lack of inversion center in noncentrosymmetric superconductors [\onlinecite{Tanaka:2009pi},\,\onlinecite{Ghosh:2010uq}].
As suggested in the first section, the Bloch band-structure topology can be apprehended via energy-band parities defined at the symmetry points ${\bf \Gamma}_{i}$. Then the momentum periodicity implies ${\cal{L}}_{\sigma \sigma'}\left({\bf k}\right)={\cal{L}}_{\sigma \sigma'}\left({\bf k}+{\bf G}\right)$ and subsequently leads to
\begin{align}
{\cal{L}}_{\sigma \sigma'}\left({\bf \Gamma}_{i}\right)=0 \,.
\end{align} 
Importantly, the Rashba spin orbit vanishes at the ${\bf \Gamma}_{i}$ points. This means that the strength of the spin-orbit can neither affect the band-structure topology, nor the existence condition of boundary Majorana quasiparticles. That is why the topological criterion introduced in the literature, namely
\begin{align}
V_{Z}^{2}>\Delta_{0}^{2}+\left(\epsilon\left({\bf \Gamma}_{i}\right) \pm \mu\right)^{2} \,,
\end{align}
does not depend on the strength of the Rashba spin-orbit coupling (see, for example, Refs.\,[\onlinecite{sau2010generic},\,\onlinecite{Alicea:2010ec},\,\onlinecite{Oreg:2010fu},\,\onlinecite{Lutchyn:2010bv},\,\onlinecite{Ghosh:2010uq},\,\onlinecite{Sato:2010kb}]). In the expression above, $\mu$ denotes the chemical potential and $\epsilon$ refers to the dispersion relation of Bloch electrons. The Rashba spin orbit plays an important role nonetheless, since it is responsible for the bulk energy gap that protects the zero-energy boundary modes, similarly to the role played by the intrinsic spin orbit in the quantum spin Hall effect [\onlinecite{Kane:2005fu,Kane:2005dz,Fu:2007oz}].

Crucially, the Hermiticity condition no longer leads to ${\cal{L}}_{\downarrow \uparrow}\left({\bf k}\right) = {\cal{L}}_{\uparrow \downarrow}^{*}\left({\bf k}\right)$ when the pattern is multiatomic. So the Rashba spin-orbit coupling is no longer antisymmetric \textit{a priori} and has no reasons to vanish at the symmetry points ${\bf \Gamma}_{i}$. This is why we expect this spin-flip process to be directly involved in the topological criterion that characterizes the existence of Majorana boundary modes in multiatomic-pattern crystals.

\subsection{Inversion-based $\Pi$ symmetry}

The BdG Hamiltonian under consideration consists of
\begin{align}\label{Total Hamiltonian}
{\cal{H}}_{0}+{\cal{H}}_{R}+{\cal{H}}_{Z}+{\cal{H}}_{S} 
&= \frac{1}{2}\sum_{{\bf k}}\psi^{\dagger}\left({\bf k}\right){\mathcal{H}}\left({\bf k}\right)\psi\left({\bf k}\right)~.
\end{align}
The multiplicative factor $1/2$ arises from the mean-field description of superconductivity. It takes into account the doubling of the degrees of freedom that is required to represent the BdG matrix ${\mathcal{H}}\left({\bf k}\right)$ in the basis of electron and hole operators.
The band structure is then fully characterized by this $8\times8$ BdG matrix that is generically written as
\begin{align}\label{Hamiltonian Matrix1}
{\cal{H}}({\bf k})=\left(\begin{array}{cc}
H({{\bf k}}) & \Delta({{\bf k}})\\
-\Delta^{*}(-{{\bf k}}) & -H^{*}({-{\bf k}}) \\
\end{array}\right)~,
\end{align}
while the explicit expression of vector $\psi$ is
\begin{align}\label{Basis Vector}
\psi^{\dagger}({\bf k})=\big(a^{\dagger}_{{\bf k} \uparrow},b^{\dagger}_{{\bf k} \uparrow},a^{\dagger}_{{\bf k} \downarrow},b^{\dagger}_{{\bf k} \downarrow},a_{-{\bf k} \uparrow},b_{-{\bf k} \uparrow},a_{-{\bf k} \downarrow},b_{-{\bf k} \downarrow} \big)~.
\end{align}
Within the mean-field description of superconductivity, the BdG matrix (\ref{Hamiltonian Matrix1}) inherently satisfies PHS as defined in Eq.\,(\ref{PHS}). The charge-conjugation operator is given here by
\begin{align}
{\mathcal{C}}=s_{0}\otimes \sigma_{0}\otimes \tau_{1} \,,
\end{align}
where $s$, $\sigma$, and $\tau$ are Pauli matrices referring to the sublattice, spin, and charge subspaces, respectively. Importantly, it satisfies ${\cal{C}}^{2}=+1$, so the system, which additionally breaks TRS, belongs to BdG class D, according to the symmetry table of Altland and Zirnbauer [\onlinecite{Altland:1997fy}]. The topology of the Bloch band structure is then characterized by a ${\mathbb{Z}}_{2}$ or ${\mathbb{Z}}$ topological invariant in one or two dimensions, respectively [\onlinecite{schnyder2008classification}].

In order to understand to what extend the sublattice structure affects the emergence of Majorana boundary modes, one then has to determine these ${\mathbb{Z}}_{2}$ or ${\mathbb{Z}}$ topological invariants. Their evaluation basically requires the knowledge of both the spectrum and the Bloch wavefunctions for all ${\bf k}$, which unfortunately implies here the diagonalization of the $8\times8$ BdG matrix $\mathcal{H}({\bf k})$. Nevertheless, it is possible to show that this matrix additionally has $\Pi$S as defined in Eq.\,(\ref{PiS}), which provides a simpler way to access these topological invariants according to the prescription given in the previous section. The definition of $\Pi$S, as well as the construction of operator $\mathcal{P}$ it relies on, are the purposes of the subsequent lines.

Let us first generically write the $4\times4$ blocks of the BdG matrix as
\begin{align}
H\left({\bf k}\right) = \left(\begin{array}{cc} K_{\uparrow \uparrow}({\bf k}) & L_{\uparrow \downarrow}({\bf k})\\L_{\downarrow \uparrow}({\bf k}) & K_{\downarrow \downarrow}({\bf k}) \\\end{array}\right) ~
\end{align}
and
\begin{align}
\Delta\left({\bf k}\right) = \left(\begin{array}{cc} 0 & D_{\uparrow \downarrow}({\bf k})\\ D_{\downarrow \uparrow}({\bf k}) & 0 \\
\end{array}\right)~.
\end{align}
The off-diagonal elements of $\Delta\left({\bf k}\right)$ are null for the discussion is limited to spin-singlet superconductivity. Block $K_{\sigma\sigma}$ describes the hopping processes, as well as the on-site chemical and Zeeman potentials in our model. Importantly, all these microscopic mechanism are invariant by inversion symmetry. In momentum space, this symmetry consists of exchanging the two sublattices A and B, and reversing the momentum ${\bf k}$ into $-{\bf k}$. It can be written as
\begin{align}\label{inversion1}
s_{1} \, K_{\sigma\sigma}({\bf k}) \, s_{1} = K_{\sigma\sigma}(-{\bf k}) ~.
\end{align}
From (\ref{Superconducting Hamiltonian}), it can be checked that block $D_{\sigma\sigma'}$, which describes spin-singlet superconductivity, satisfies a similar relation
\begin{align}
\label{inversion2}
s_{1} \, D_{\sigma\sigma'}({\bf k}) \, s_{1} = D_{\sigma\sigma'}(-{\bf k})~.
\end{align}
The Rashba spin-orbit does not break the TRS. As mentioned earlier, this leads to $L_{\downarrow \uparrow}({\bf k})=-L_{\uparrow \downarrow}^{*}(-{\bf k})$. The time-reversal invariance implies, along with Hermiticity, that
\begin{align}
\label{inversion3}
s_{1} \, L_{\sigma \sigma'}({\bf k}) \, s_{1} = -L_{\sigma \sigma'}(-{\bf k})~.
\end{align}
As a result of Eqs. (\ref{inversion1}), (\ref{inversion2}) and (\ref{inversion3}), BdG matrix $\mathcal{H}$ has $\Pi$S as defined in Eq.\,(\ref{PiS}), that is
\begin{align}
\label{PiS 2}
\mathcal{P}\,\mathcal{H}({\bf k})\,\mathcal{P}^{-1} &= \mathcal{H}(-{\bf k}) ~~~~~~\text{with}~~~~~~
{\cal P}=s_{1}\otimes \sigma_{3}\otimes\tau_{3} \,.
\end{align}
Note that this relation looks like Eq.\,(\ref{inversion1}) that defines inversion symmetry. This is the reason why the paritylike relation above is referred to as $\Pi$ symmetry throughout this paper.

The $\Pi$ basis has been defined as the basis in which operator ${\cal P}$ has the diagonal representation ${\tilde{\cal P}} = \sigma_{0} \otimes s_{0} \otimes \tau_{3}$. For operator $\mathcal{P}$ given in Eq.\,(\ref{PiS 2}), the $\Pi$-basis is obtained via the unitary operator $U$ defined as
\begin{align}
2U=\sigma_0\otimes s_0\otimes(\tau_0+\tau_1) +\sigma_1\otimes s_3\otimes(\tau_0-\tau_1) \,.
\end{align}
Interestingly, $\Pi$S (\ref{PiS 2}) requires the BdG matrix to satisfy
\begin{align}
\label{BdG Matrix 1}
\tilde{\cal{H}}({\bf k})=\left(\begin{array}{cc}
\tilde{H}\left({{\bf k}}\right) & \tilde{\Delta}\left({{\bf k}}\right)\\
-\tilde{\Delta}^{*}({-{\bf k}}) & -\tilde{H}^{*}({-{\bf k}}) \\
\end{array}\right) \,,
\end{align}
with $\tilde{H}\left({{\bf k}}\right)=\tilde{H}\left({-{\bf k}}\right)$ and $\tilde{\Delta}\left({{\bf k}}\right)=-\tilde{\Delta}\left(-{{\bf k}}\right)$. This conceptually means that $\mathcal{H}$ can be mapped onto an effective band structure that describes an odd-parity superconductor where inversion symmetry would be associated to operator $\tilde{\mathcal{P}}$ [\onlinecite{Sato:2009fe},\,\onlinecite{Sato:2010lh}].

\begin{figure}[t]
\includegraphics[trim = 0mm 80mm 0mm 80mm, clip, width=8cm]{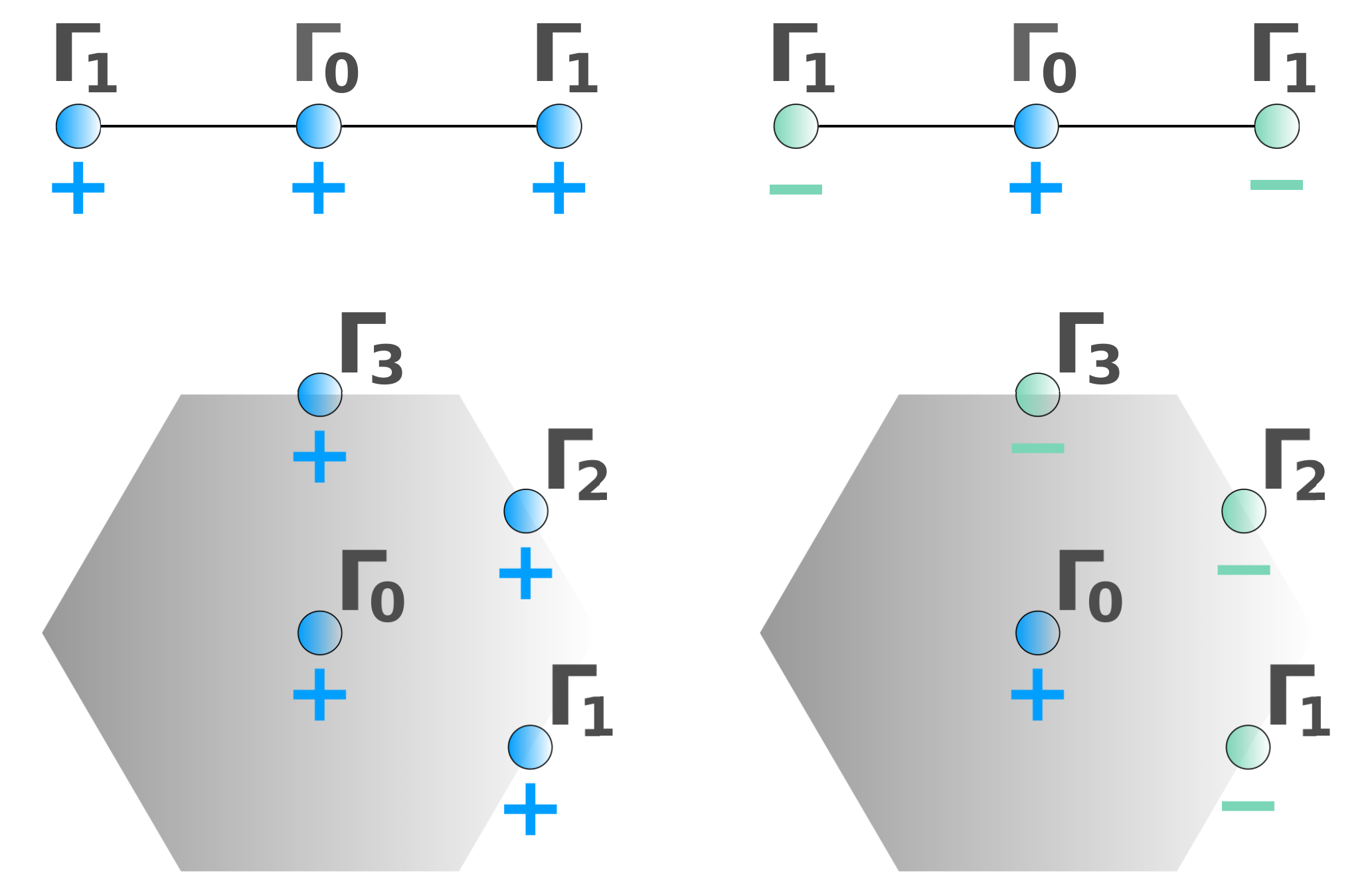}
\caption{\small (Color online) Illustrations of the parity products $\delta_{i}=\pm1$ at the symmetric points ${\Gamma_{i}}$ within one dimension (top) and two dimension (bottom) BZ. When a system is associated to the trivial configuration depicted in the left-hand column, it necessarily has to undergo some band inversions to reach the topological configuration of the right-hand column.}
\label{Parity Products}
\end{figure}

\begin{figure*}[p]
\centering
$\begin{array}{ccc}
~~~~~~~\includegraphics[trim = 25mm 0mm 30mm 0mm, clip, width=4.7cm]{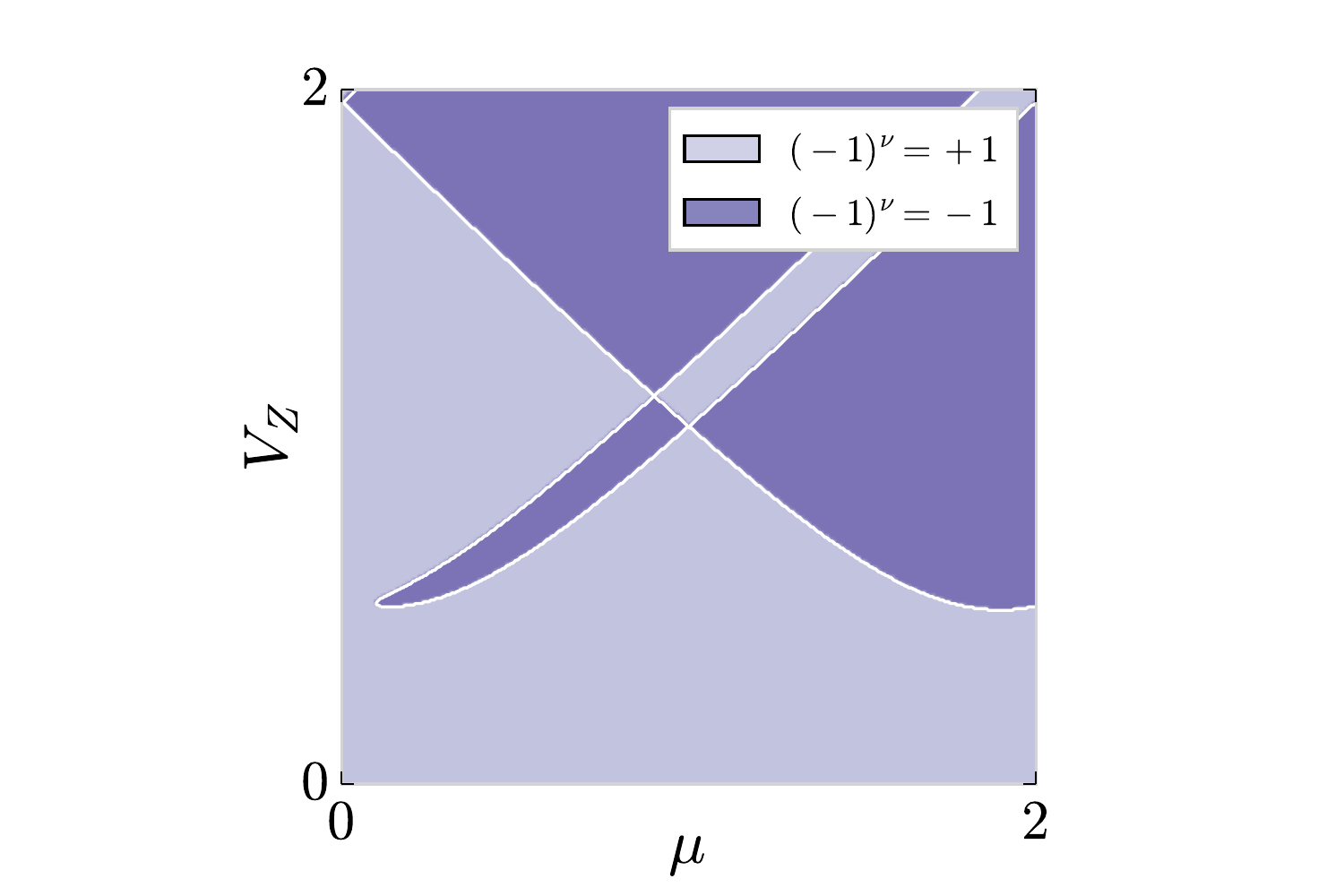}~~~~~~~~&
~~~~~~~~\includegraphics[trim = 25mm 0mm 30mm 0mm, clip, width=4.7cm]{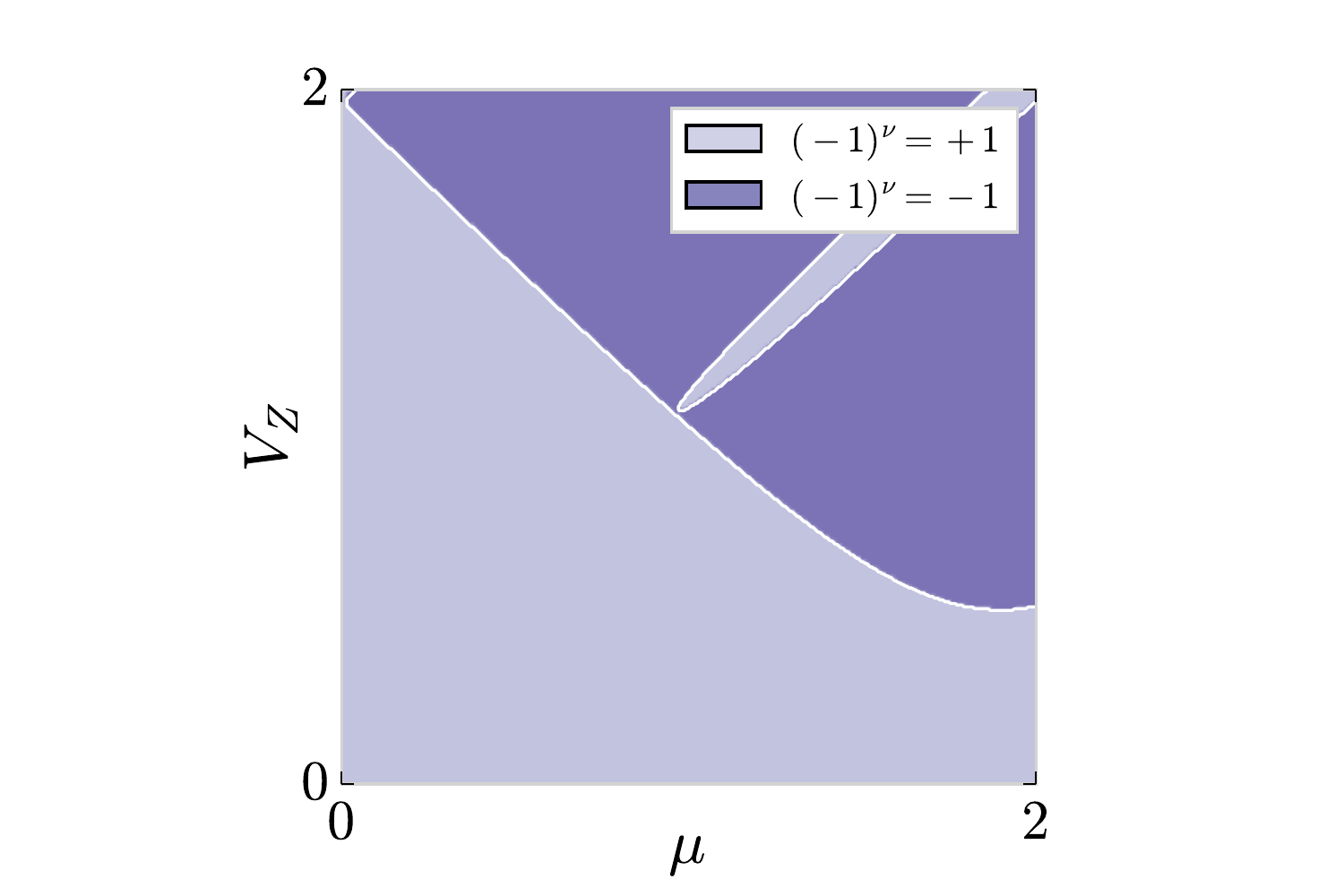}~~~~~~~~&
~~~~~~~~\includegraphics[trim = 25mm 0mm 30mm 0mm, clip, width=4.7cm]{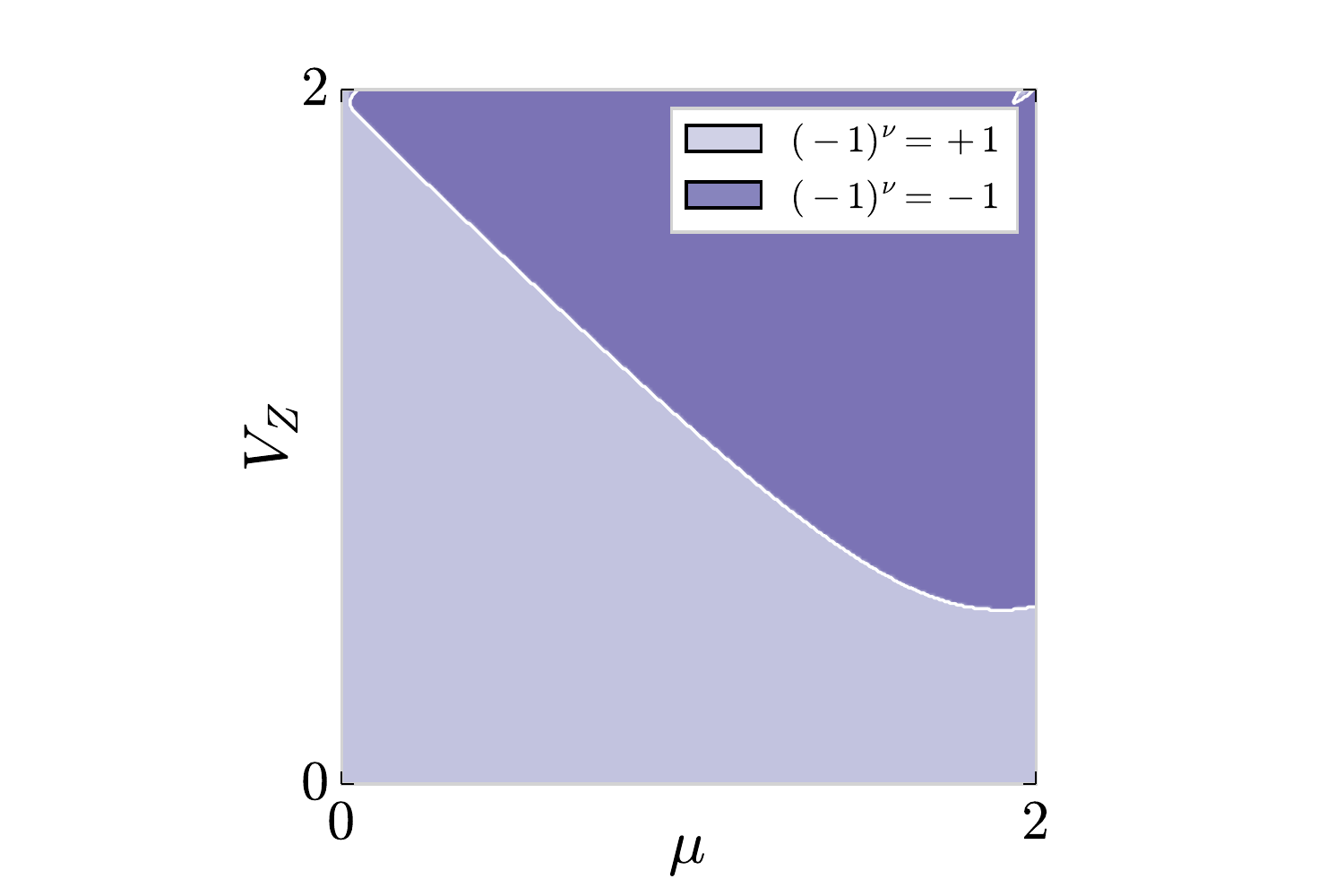}~~~~~~~~\\
~~~~~~~\includegraphics[trim = 25mm 0mm 30mm 0mm, clip, width=4.7cm]{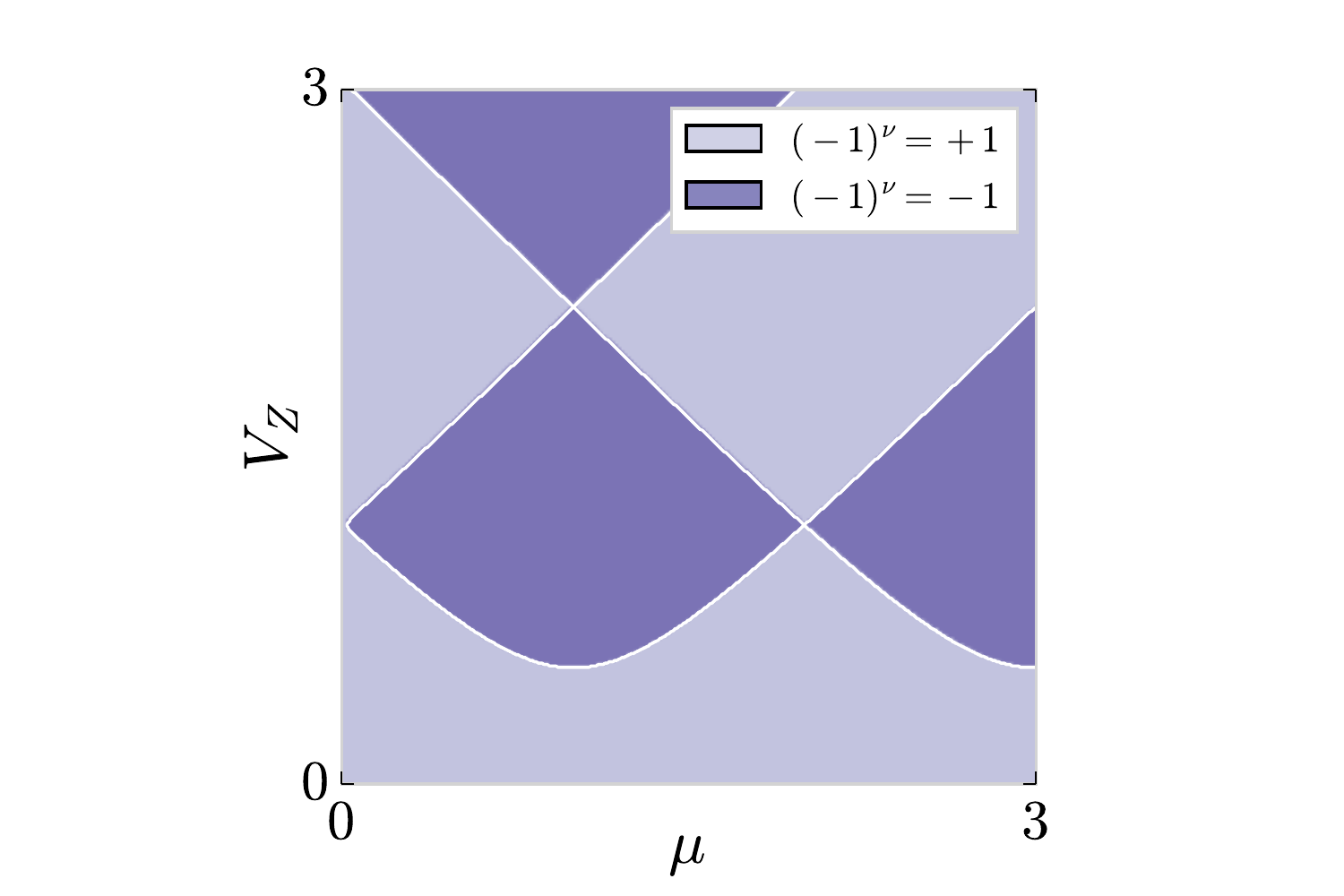}~~~~~~~~&
~~~~~~~~\includegraphics[trim = 25mm 0mm 30mm 0mm, clip, width=4.7cm]{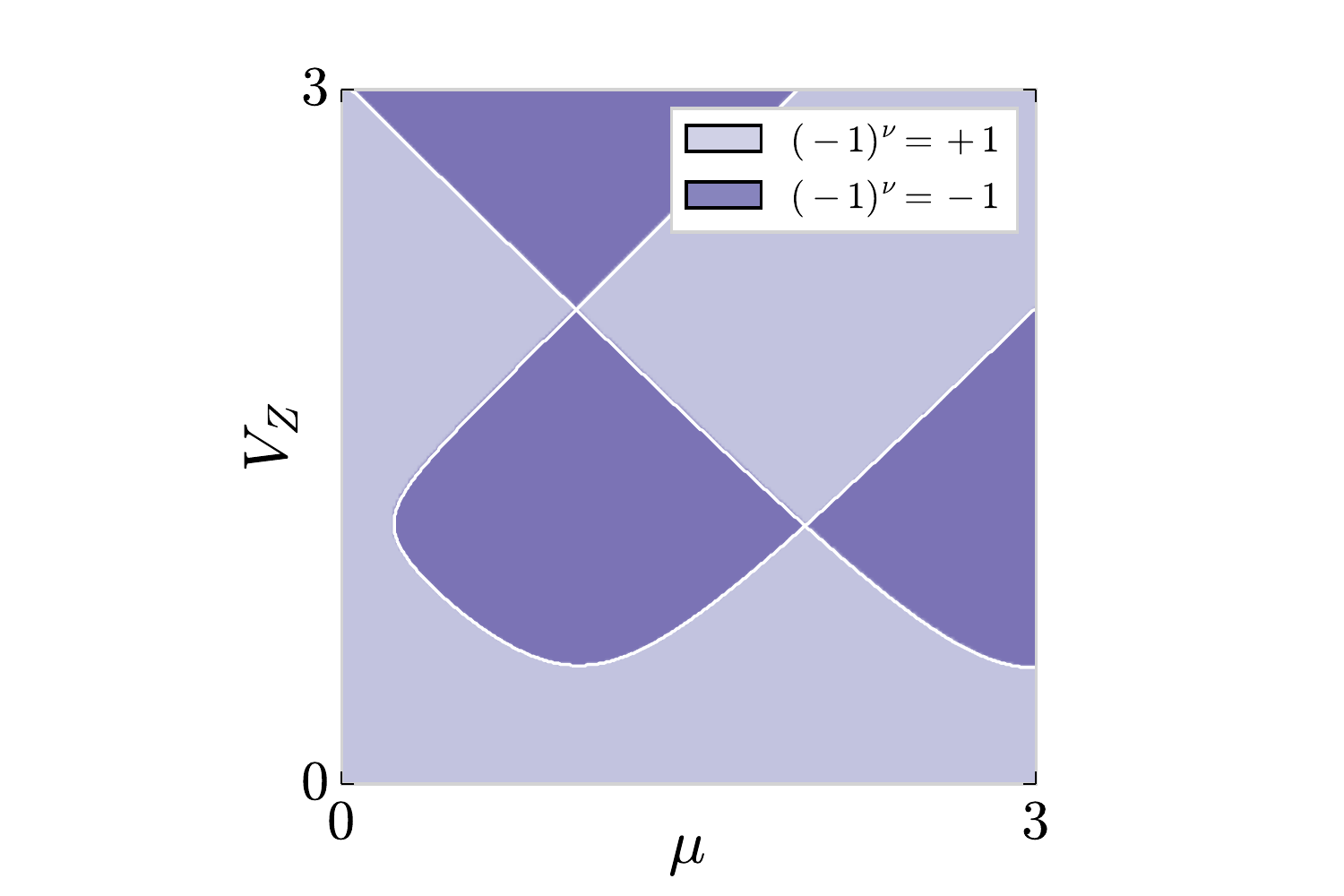}~~~~~~~~&
~~~~~~~~\includegraphics[trim = 25mm 0mm 30mm 0mm, clip, width=4.7cm]{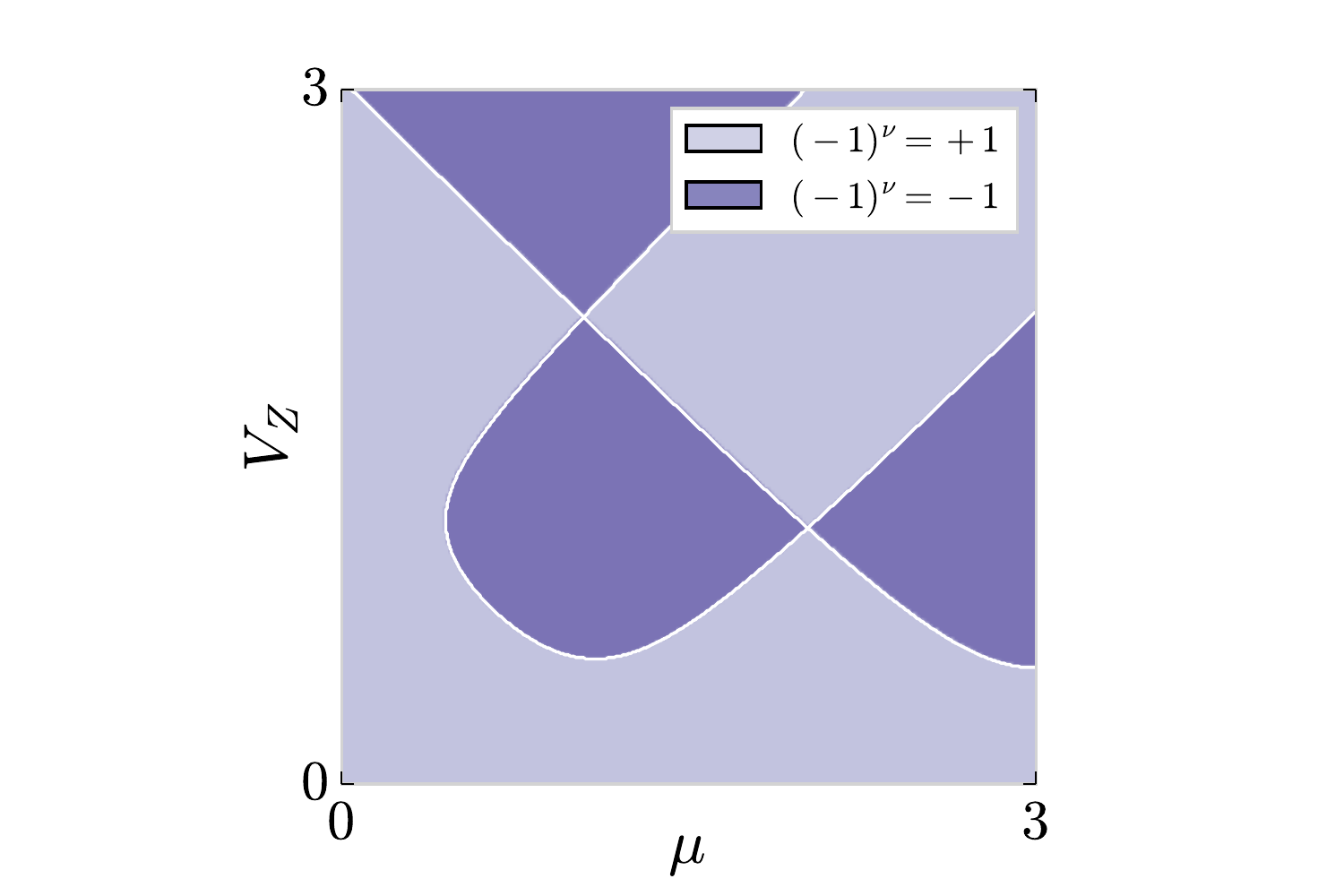}~~~~~~~~\\
~~~~~~~\includegraphics[trim = 25mm 0mm 30mm 0mm, clip, width=4.7cm]{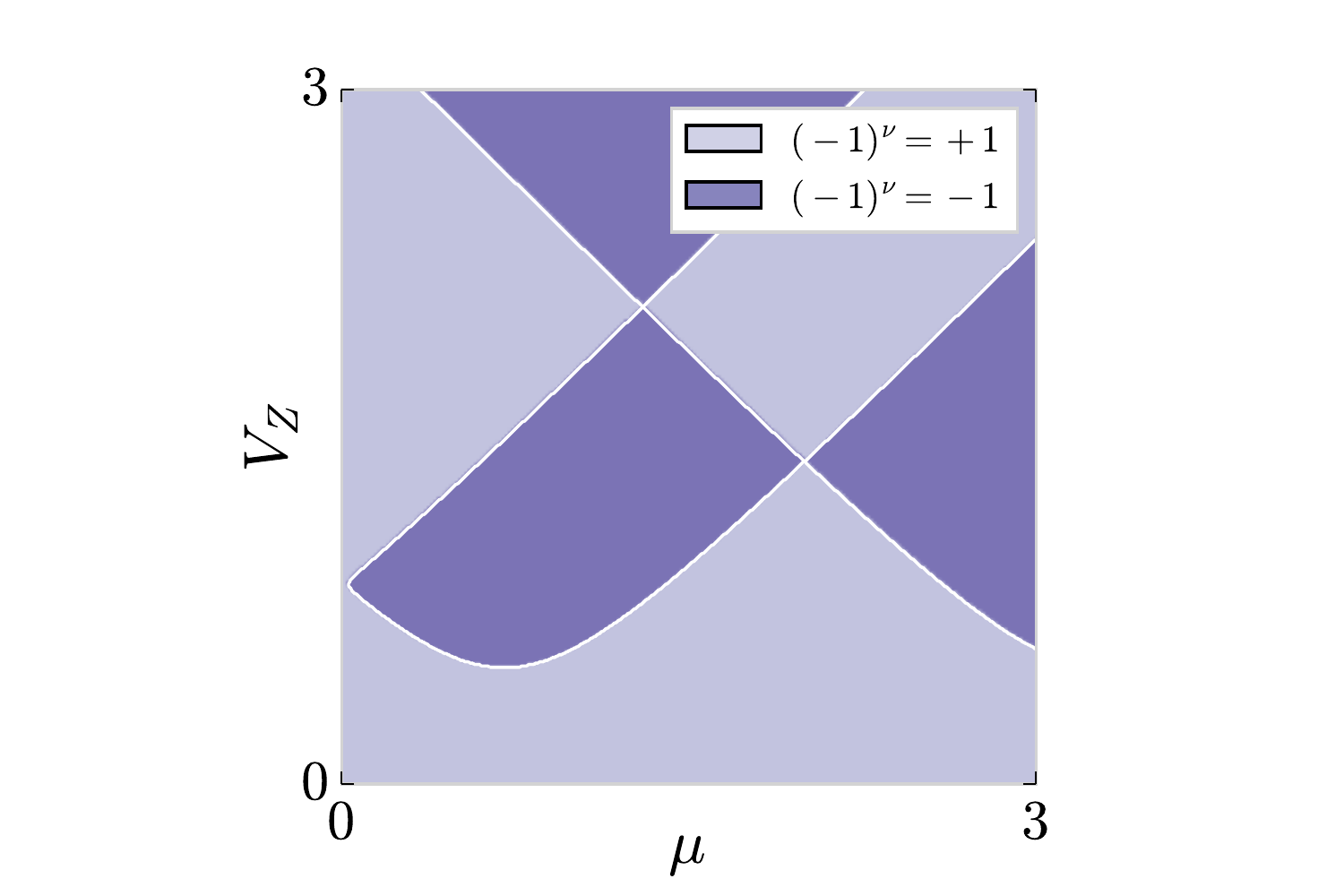}~~~~~~~~&
~~~~~~~~\includegraphics[trim = 25mm 0mm 30mm 0mm, clip, width=4.7cm]{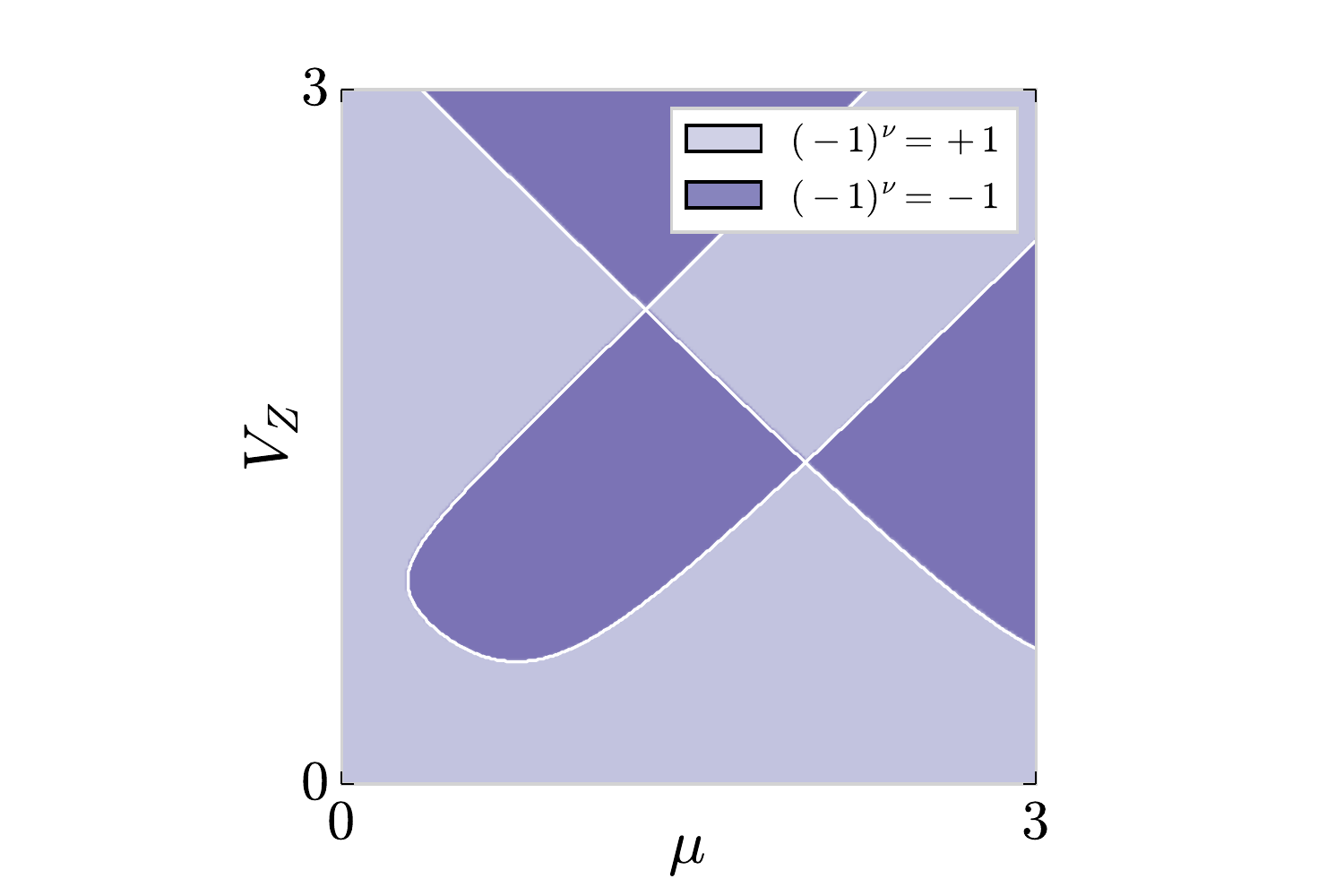}~~~~~~~~&
~~~~~~~~\includegraphics[trim = 25mm 0mm 30mm 0mm, clip, width=4.7cm]{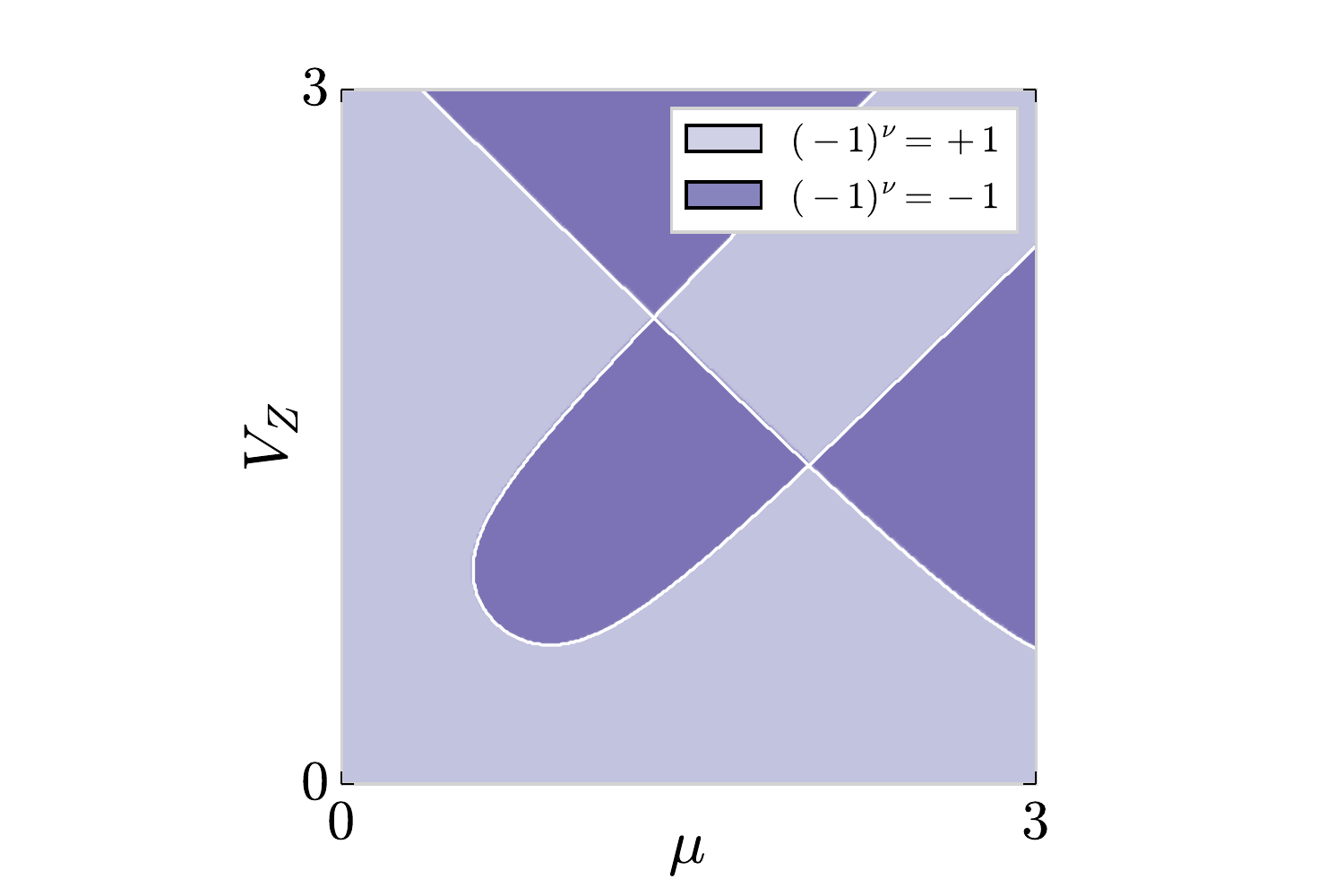}~~~~~~~~\\
~~~~~~~\includegraphics[trim = 25mm 0mm 30mm 0mm, clip, width=4.7cm]{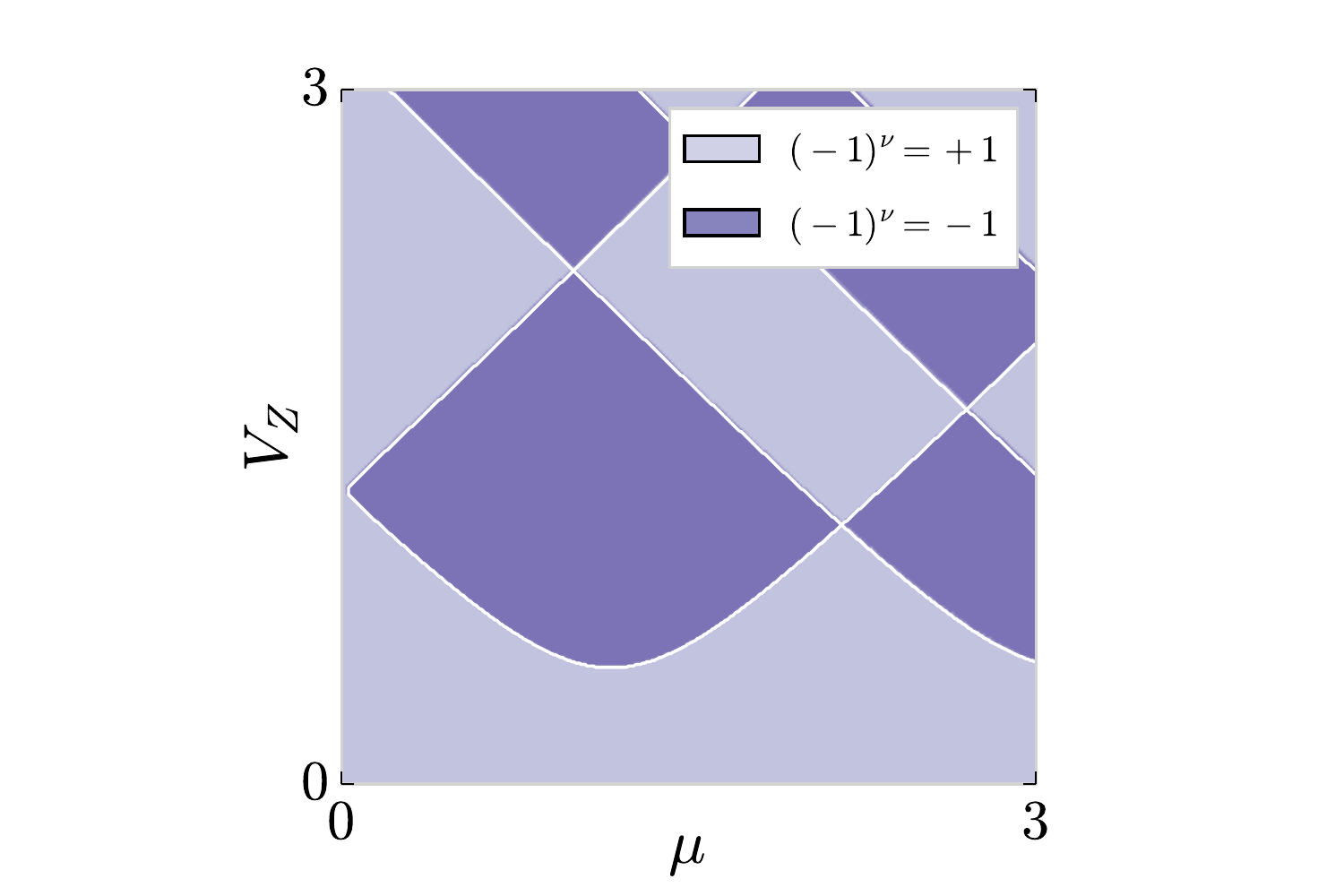}~~~~~~~~&
~~~~~~~~\includegraphics[trim = 25mm 0mm 30mm 0mm, clip, width=4.7cm]{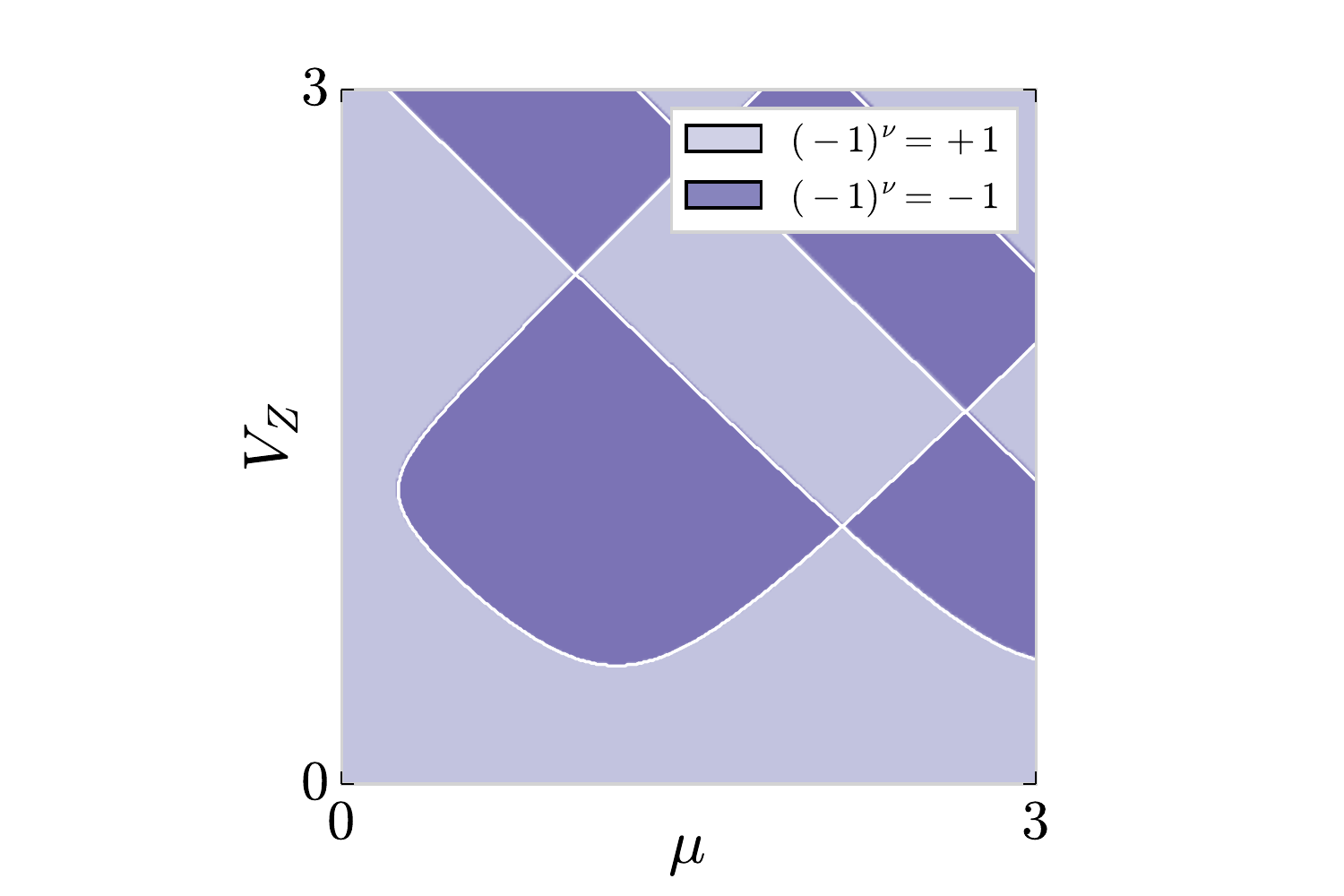}~~~~~~~~&
~~~~~~~~\includegraphics[trim = 25mm 0mm 30mm 0mm, clip, width=4.7cm]{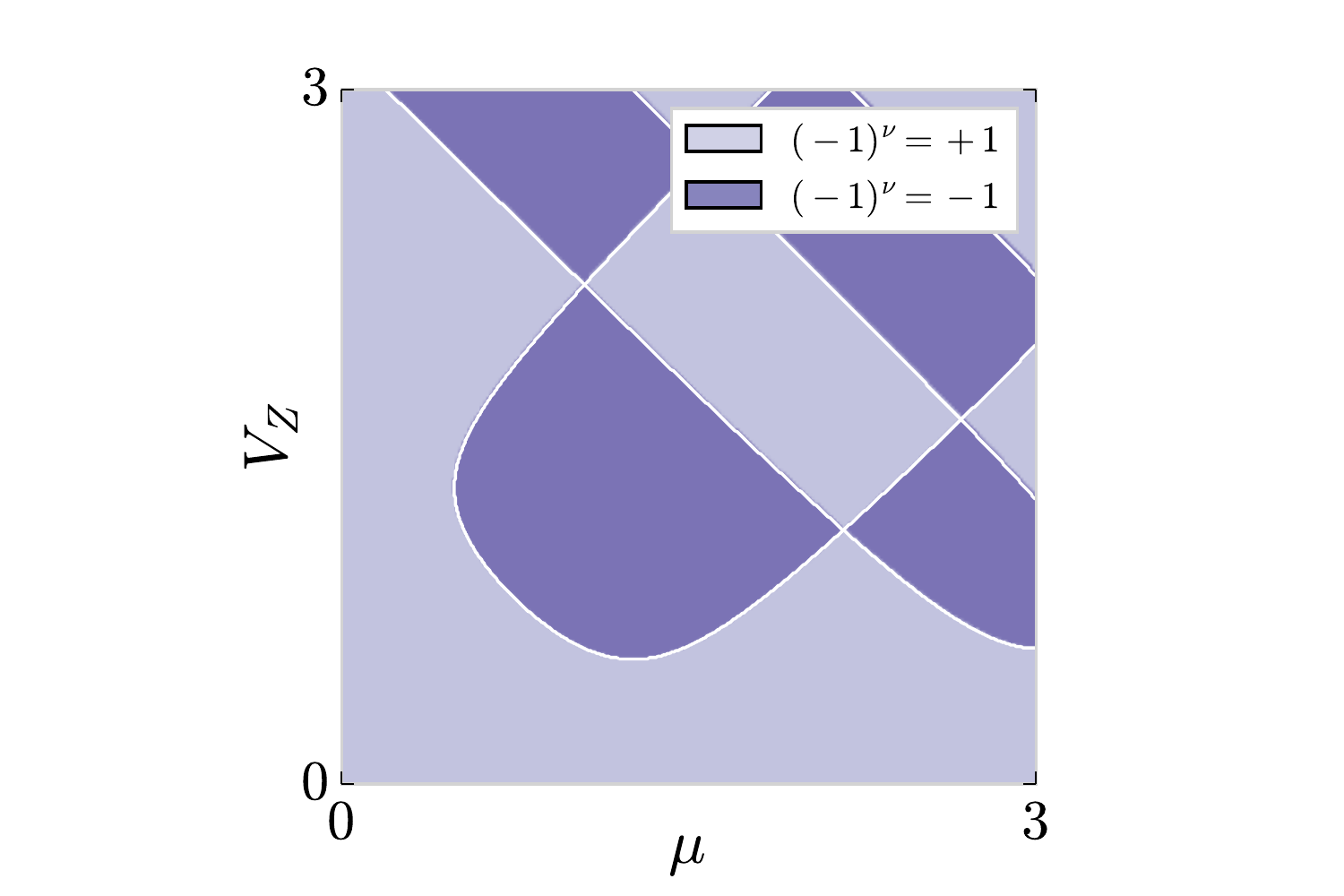}~~~~~~~~\\
\end{array}$
\caption{\small (Color online) Topological phase diagrams for the dimerized Peierls chain ($\alpha=0.9$), graphene ($\alpha=1.0$), stretched graphene ($\alpha=1.3$), and phosphorene ($\alpha=3.0$ and $V_{E}=1.0$) from top to bottom, respectively. Light (dark) purple refers to the phase characterized by $(-1)^{\nu}=+1(-1)$. The columns correspond to $\lambda=0.01$, $\lambda=0.10$, and $\lambda=0.20$ from left to right, respectively. Spin-singlet pairings have been chosen as $\Delta_{0}=0.50$ and $\Delta_{1}=0.00$ for all plots. Energy is given in units of the nearest-neighbor hopping amplitude $t$.}
\label{Zeeman Phase Diagram}
\end{figure*}

\begin{figure*}[p]
\centering
$\begin{array}{ccc}
~~~~~~~\includegraphics[trim = 25mm 0mm 30mm 0mm, clip, width=4.7cm]{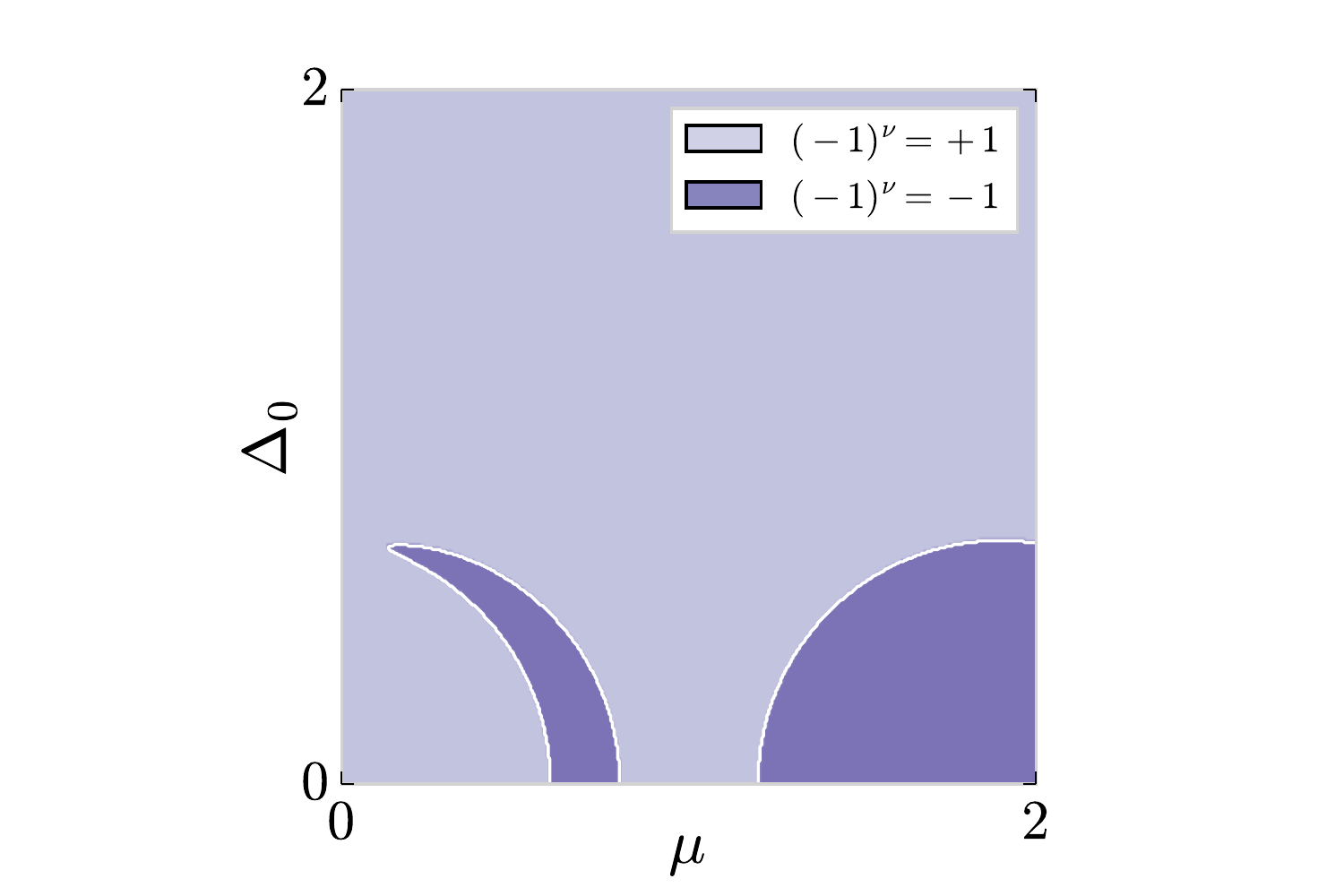}~~~~~~~~&
~~~~~~~~\includegraphics[trim = 25mm 0mm 30mm 0mm, clip, width=4.7cm]{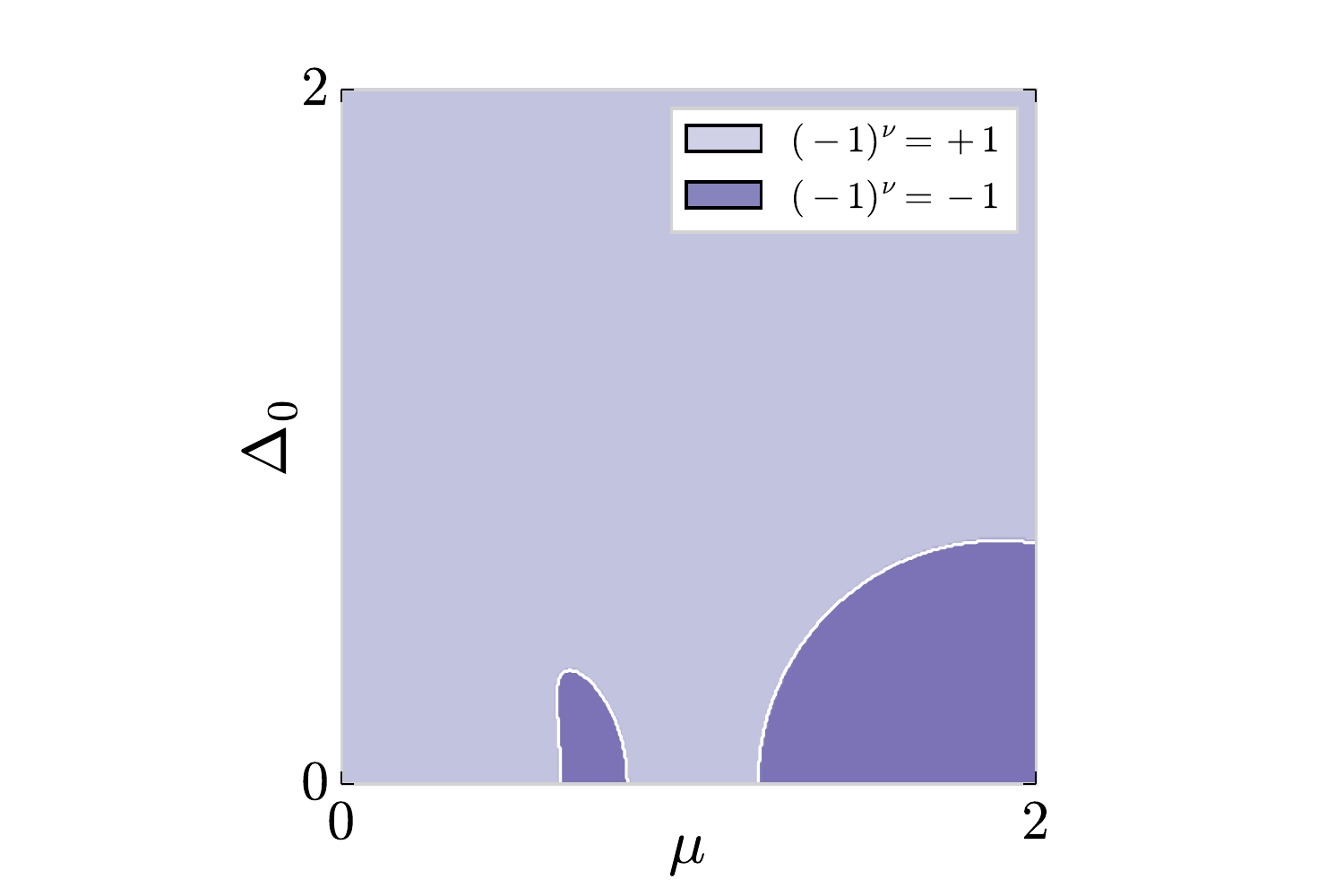}~~~~~~~~&
~~~~~~~~\includegraphics[trim = 25mm 0mm 30mm 0mm, clip, width=4.7cm]{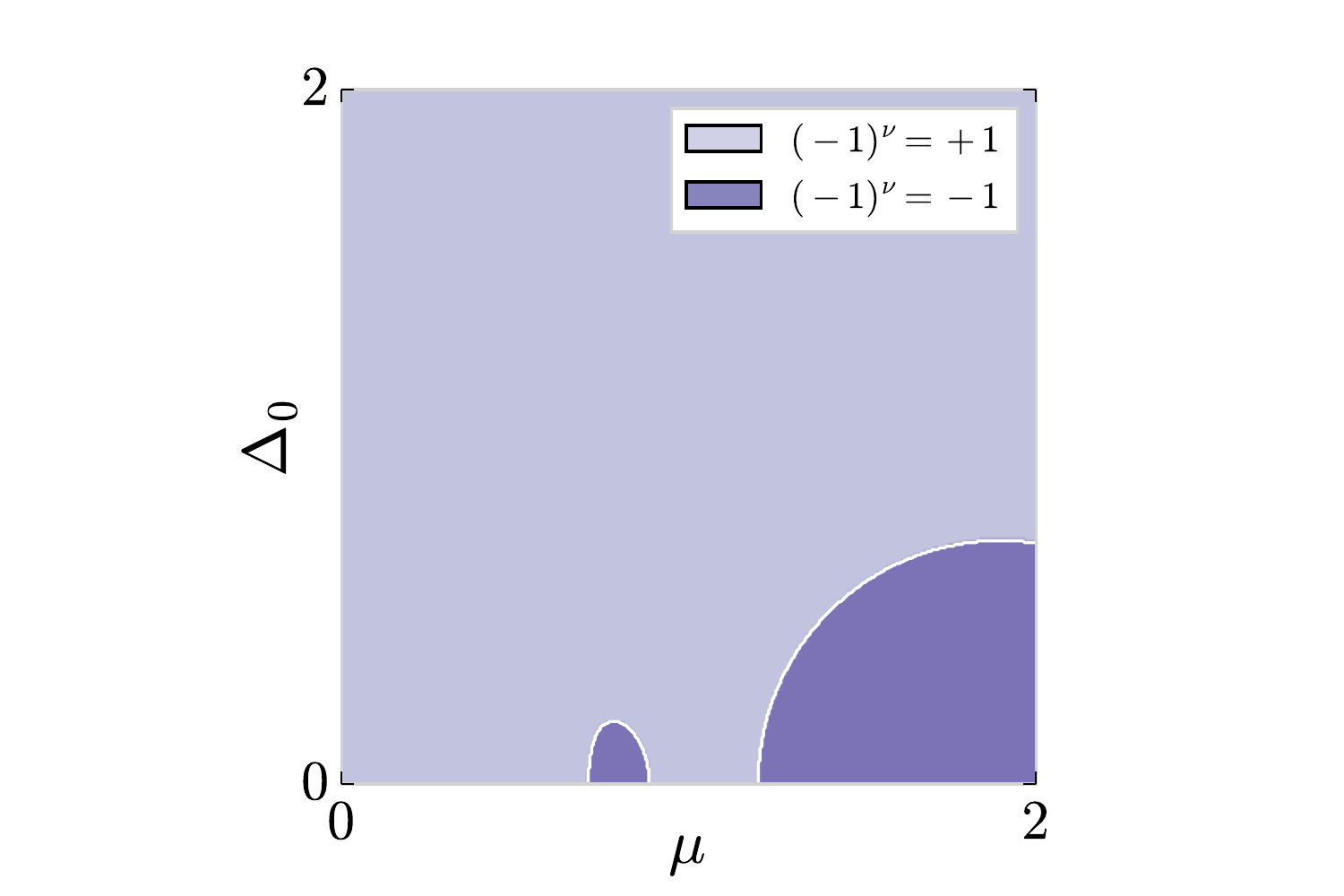}~~~~~~~~\\
~~~~~~~\includegraphics[trim = 25mm 0mm 30mm 0mm, clip, width=4.7cm]{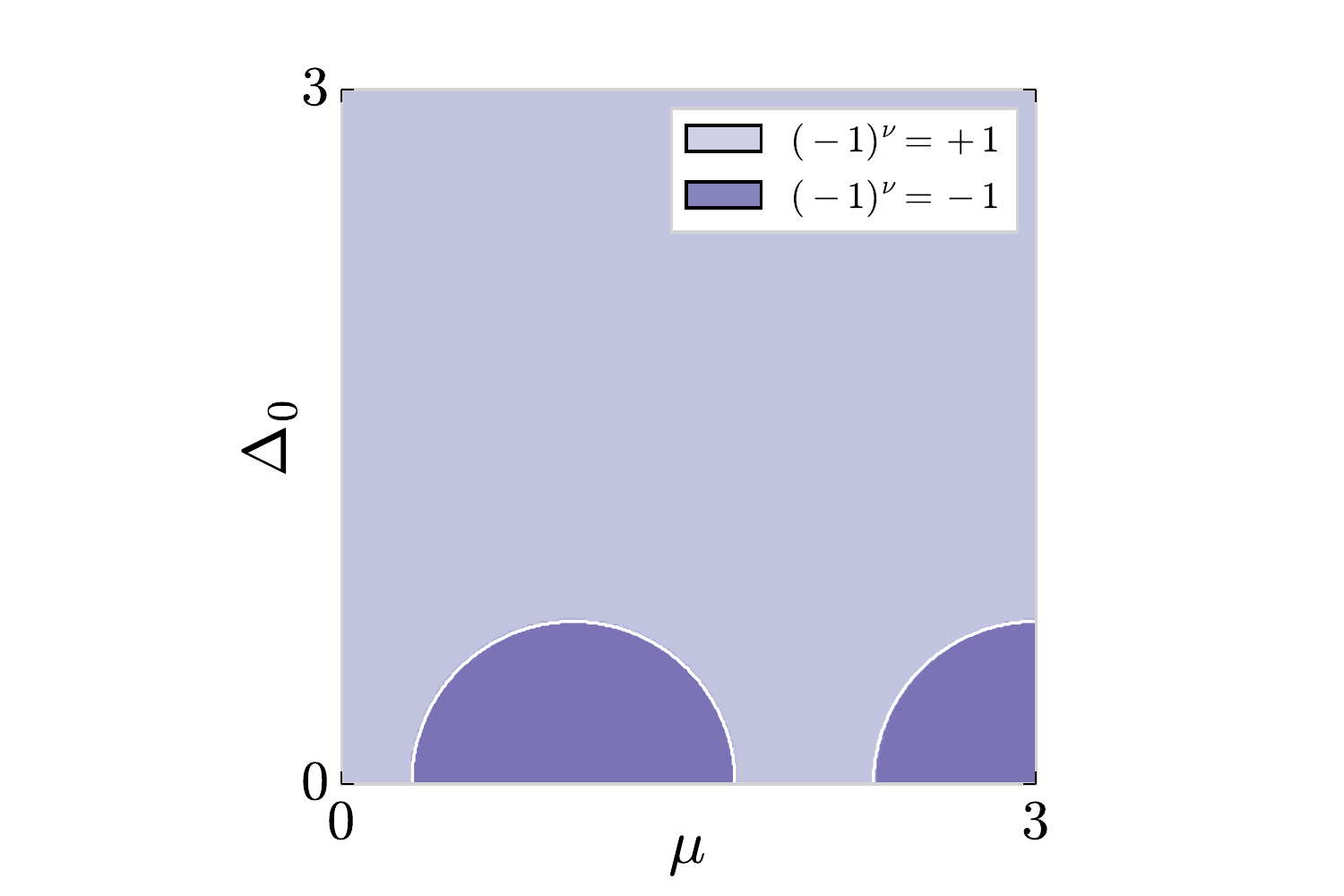}~~~~~~~~&
~~~~~~~~\includegraphics[trim = 25mm 0mm 30mm 0mm, clip, width=4.7cm]{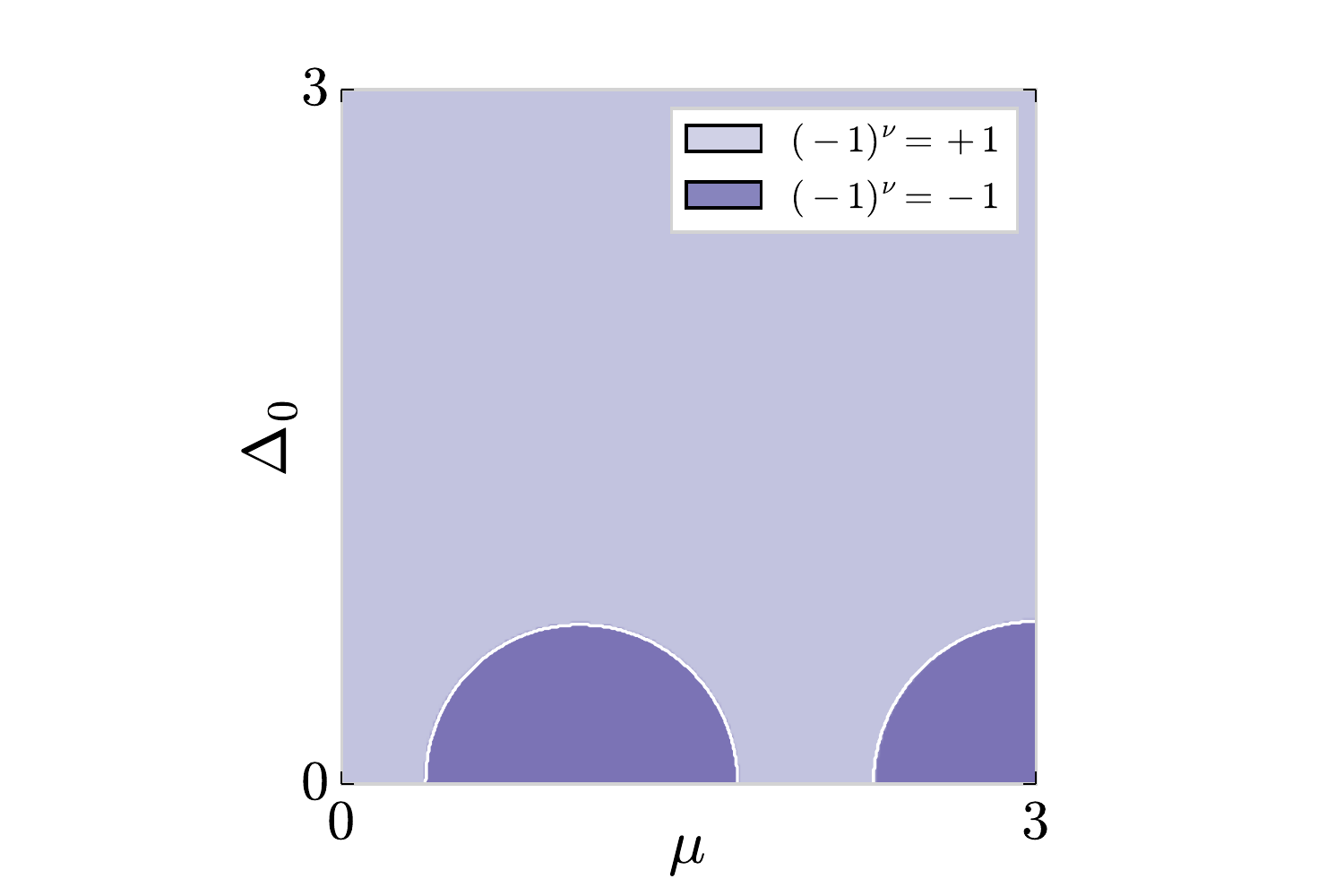}~~~~~~~~&
~~~~~~~~\includegraphics[trim = 25mm 0mm 30mm 0mm, clip, width=4.7cm]{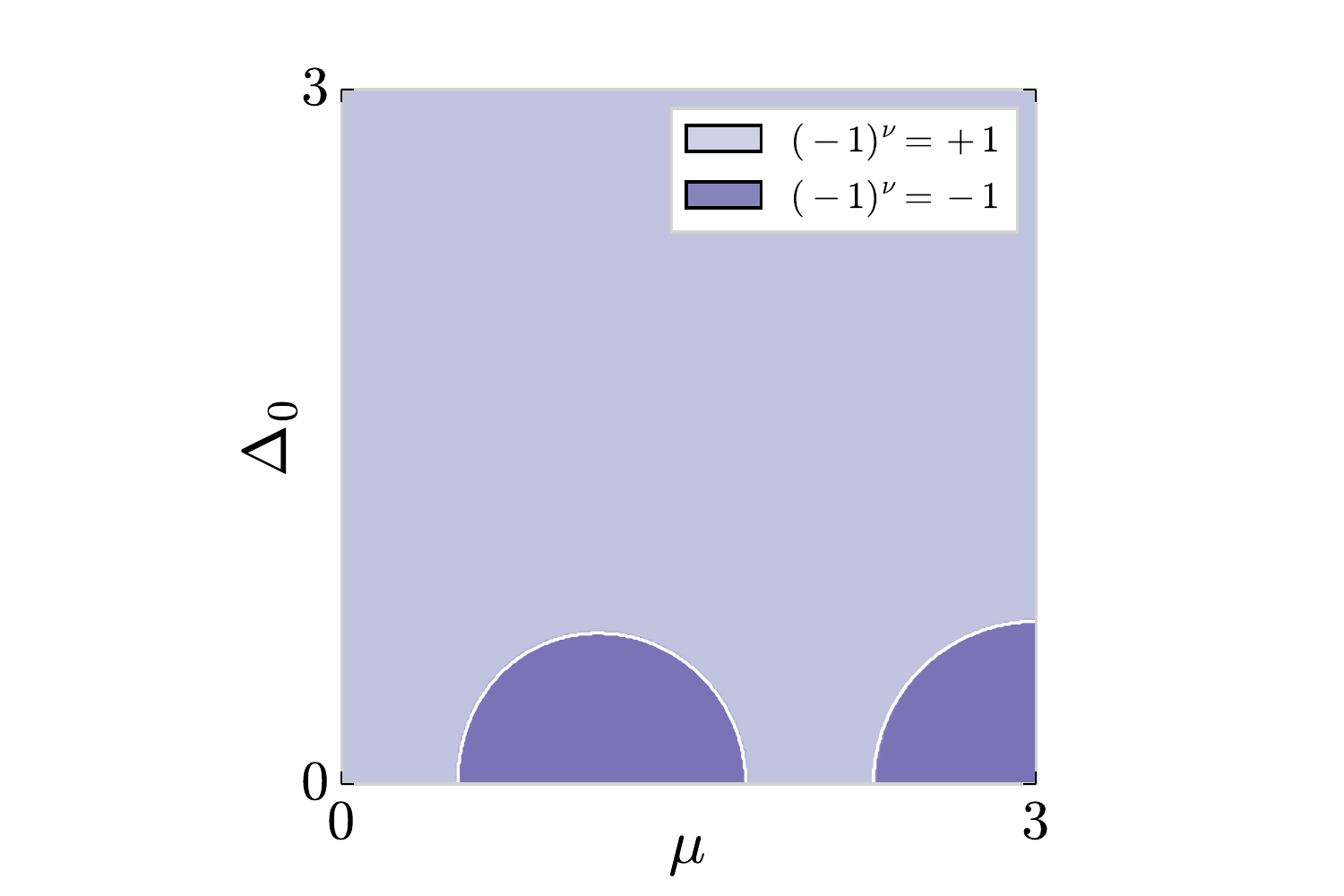}~~~~~~~~\\
~~~~~~~\includegraphics[trim = 25mm 0mm 30mm 0mm, clip, width=4.7cm]{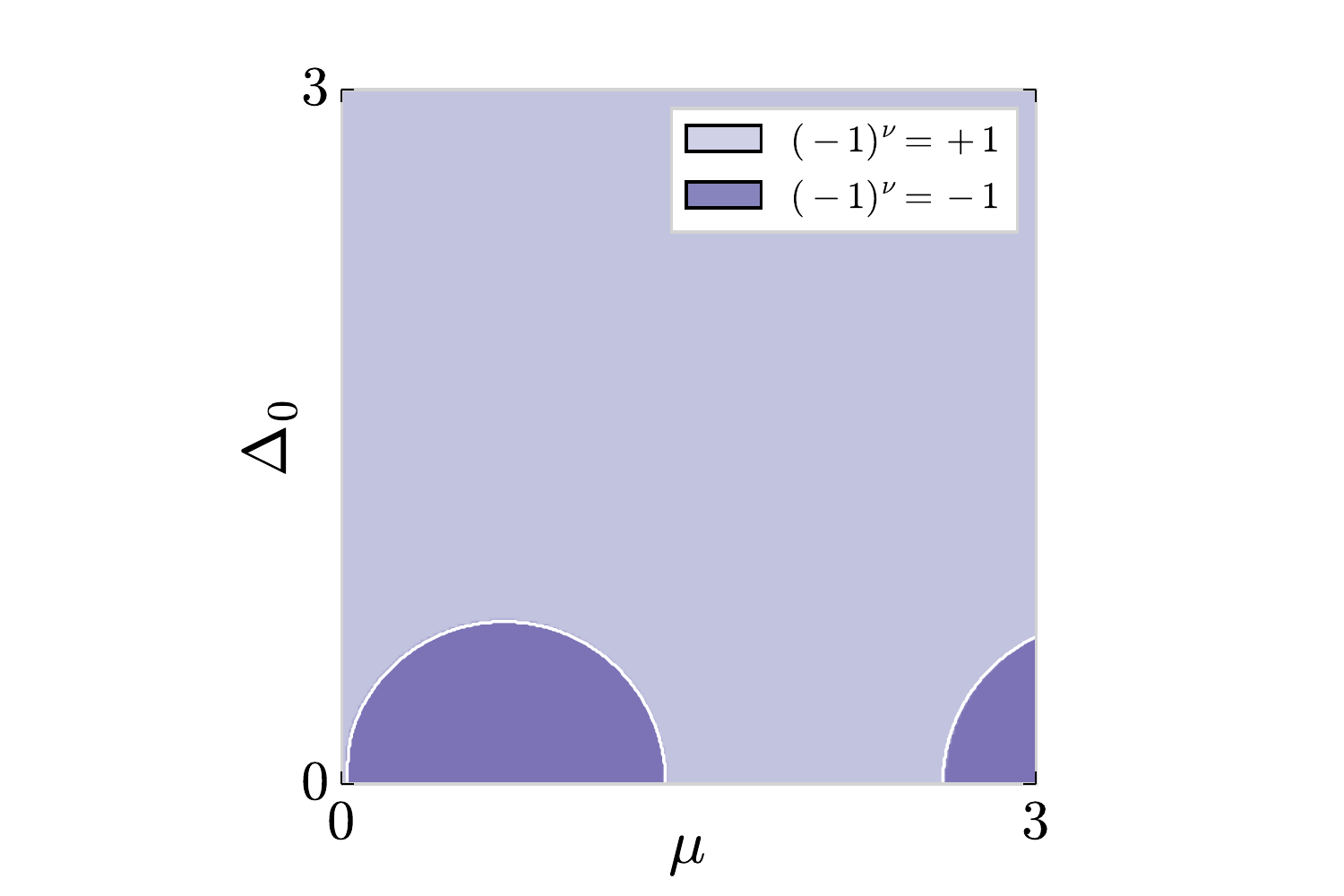}~~~~~~~~&
~~~~~~~~\includegraphics[trim = 25mm 0mm 30mm 0mm, clip, width=4.7cm]{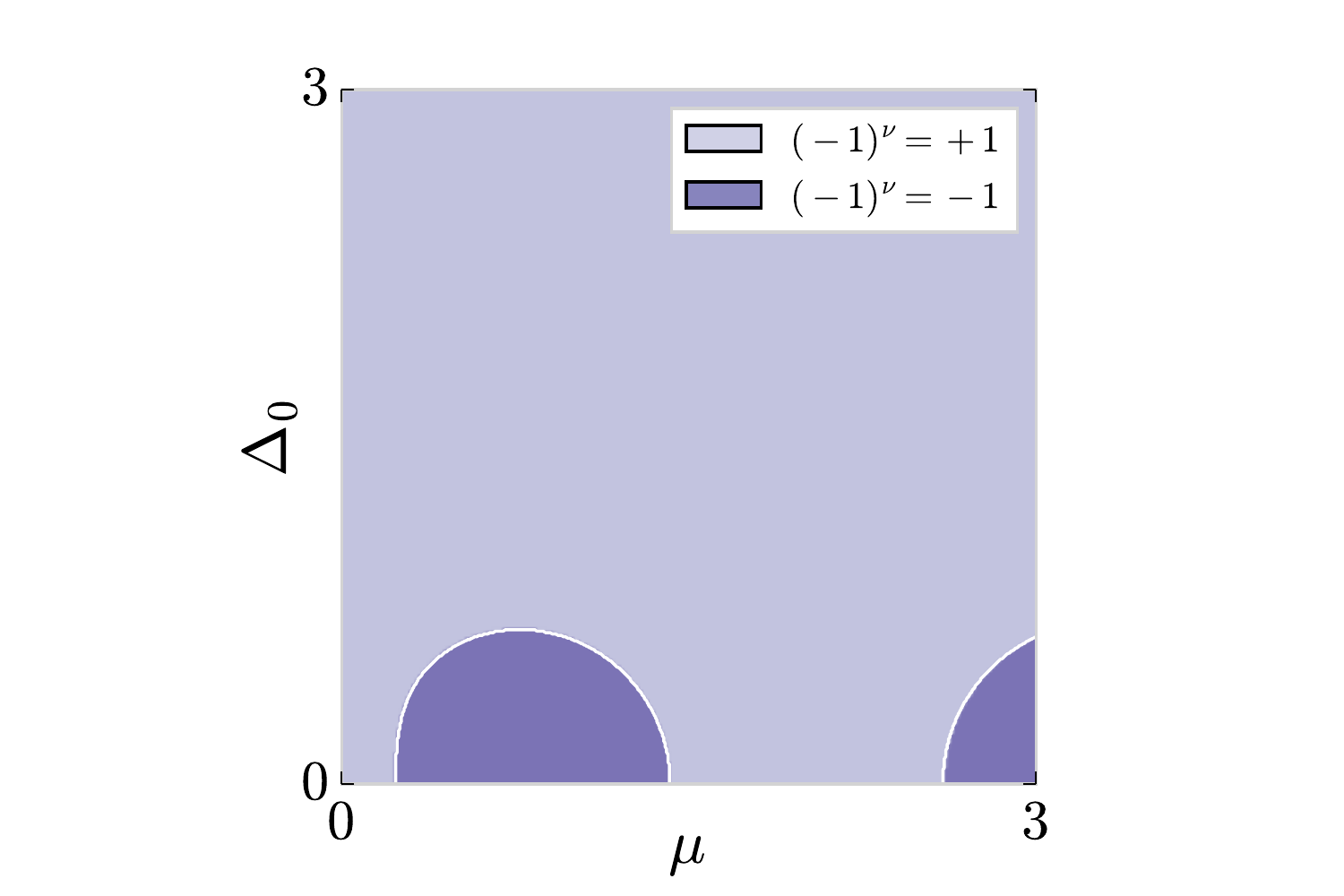}~~~~~~~~&
~~~~~~~~\includegraphics[trim = 25mm 0mm 30mm 0mm, clip, width=4.7cm]{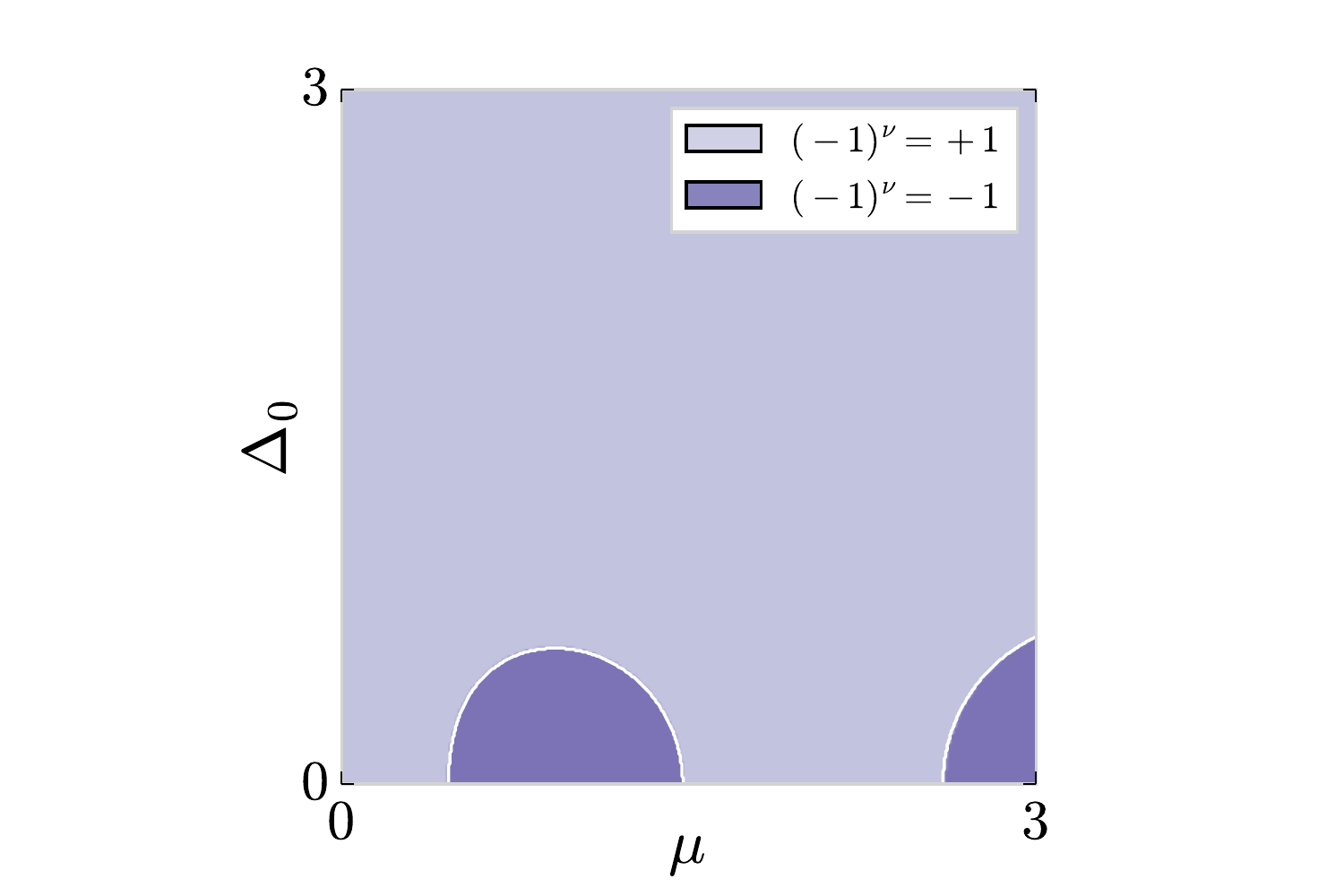}~~~~~~~~\\
~~~~~~~\includegraphics[trim = 25mm 0mm 30mm 0mm, clip, width=4.7cm]{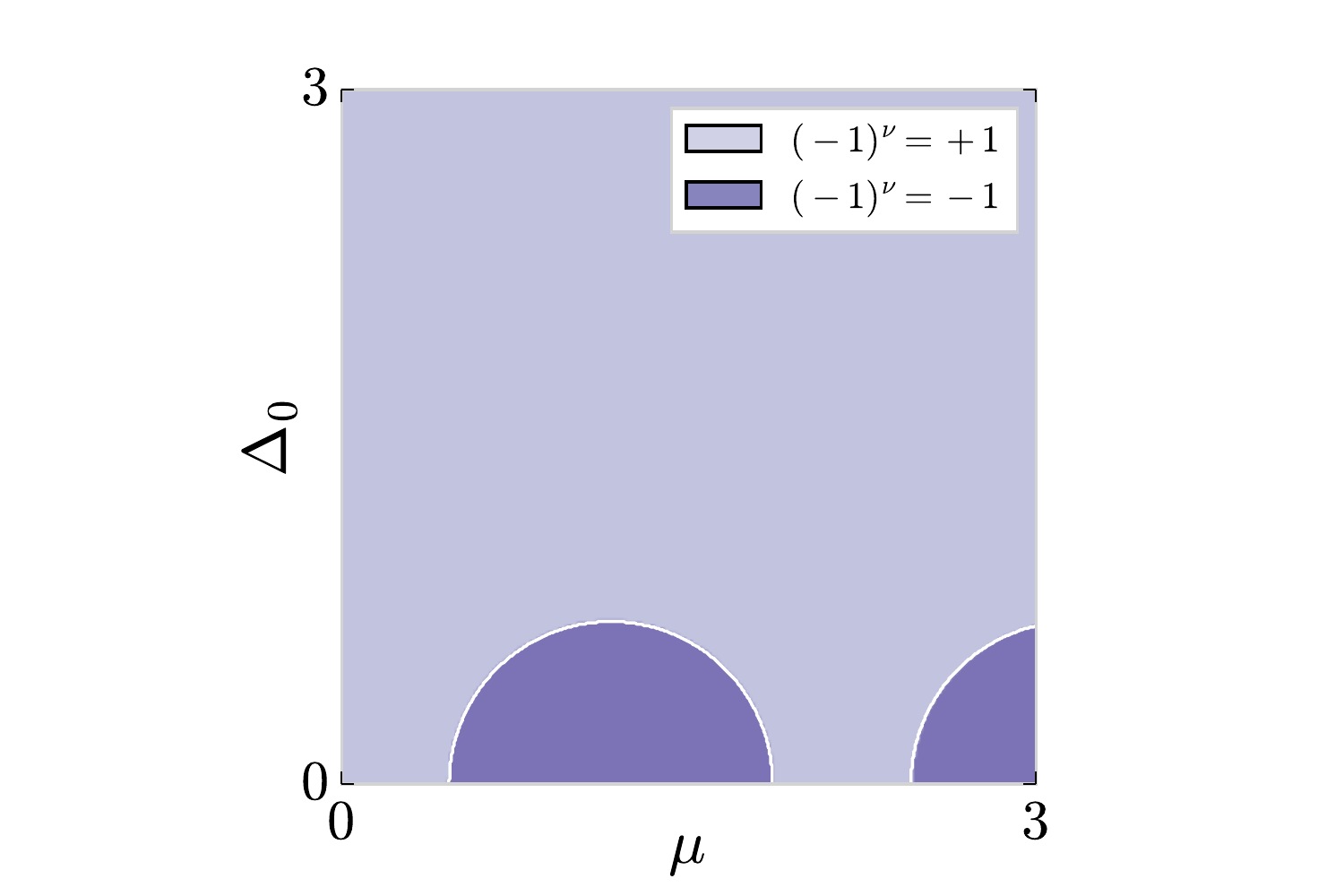}~~~~~~~~&
~~~~~~~~\includegraphics[trim = 25mm 0mm 30mm 0mm, clip, width=4.7cm]{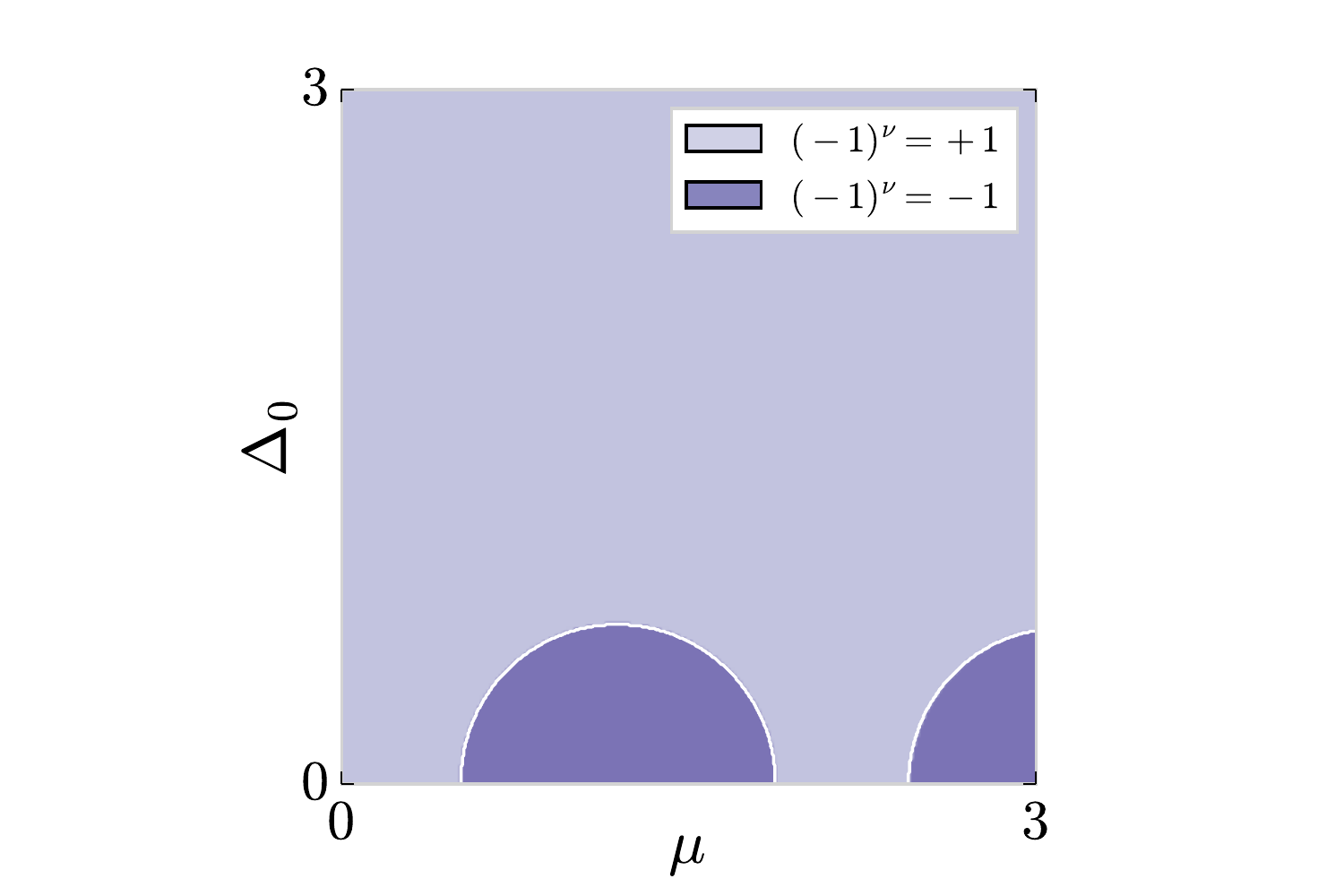}~~~~~~~~&
~~~~~~~~\includegraphics[trim = 25mm 0mm 30mm 0mm, clip, width=4.7cm]{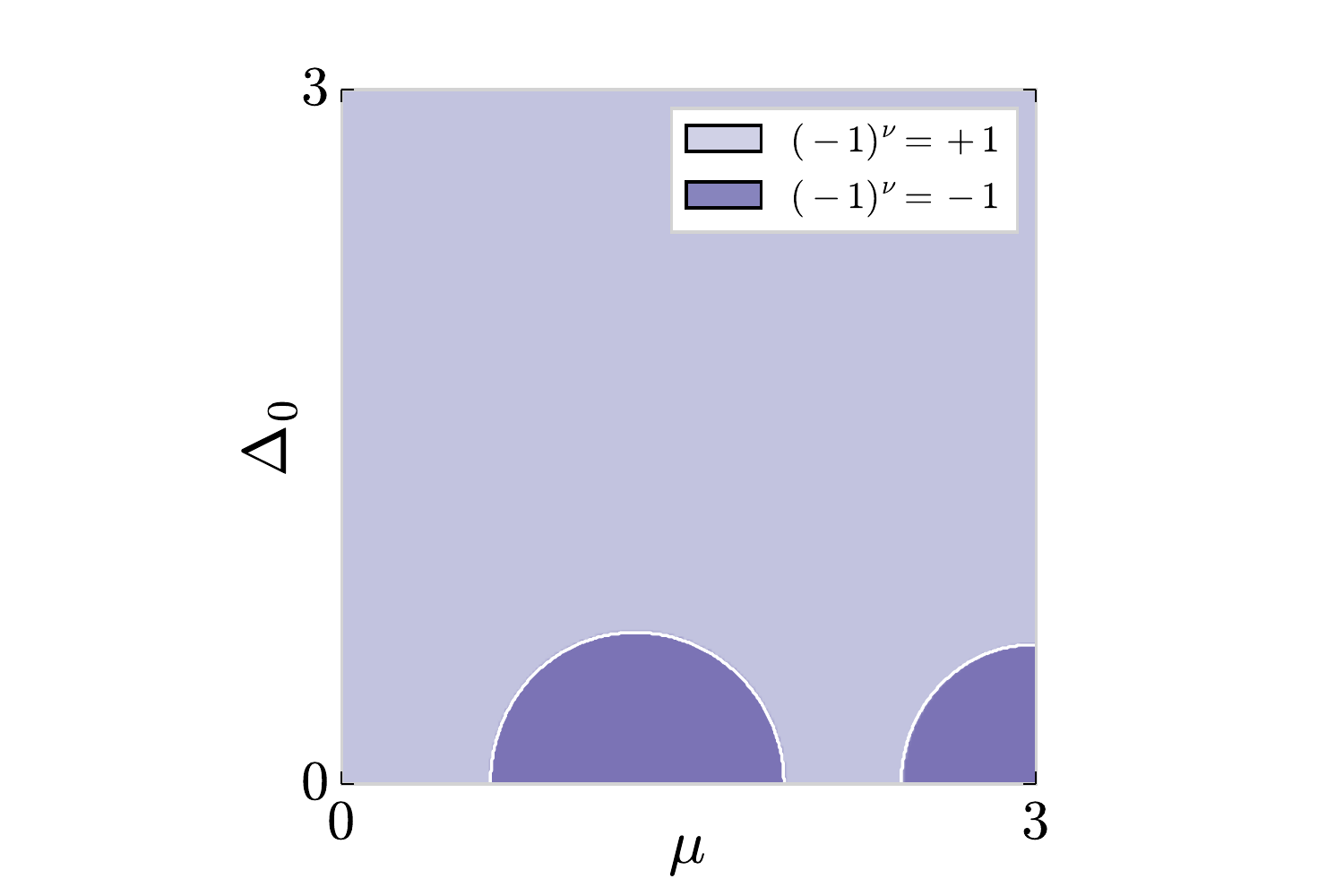}~~~~~~~~\\
\end{array}$
\caption{\small (Color online) Topological phase diagrams for the dimerized Peierls chain ($\alpha=0.9$), graphene ($\alpha=1.0$), stretched graphene ($\alpha=1.3$), and phosphorene ($\alpha=3.0$ and $V_{E}=1.0$) from top to bottom, respectively. Light (dark) purple refers to the phase characterized by $(-1)^{\nu}=+1(-1)$. The columns correspond to $\lambda=0.01$, $\lambda=0.10$, and $\lambda=0.20$ from left to right, respectively. The Zeeman potential has been chosen as $V_{Z}=0.7$ and $\Delta_{1}=0$ for all plots. Energy is given in units of the nearest-neighbor hopping amplitude $t$.}
\label{Supra Phase Diagram}
\end{figure*}

\subsection{Band inversion criteria}
Equation\,(\ref{Parity Product 3}) provides a simple criterion to characterize topological band inversions, as the ones illustrated in Fig.\,\ref{Parity Products}. It relies on the following determinant:
\begin{align}
\label{Determinant}
\Det \tilde{H}({\bf \Gamma}_i)=V_{z}^{4}-2 \, \alpha V_{z}^{2}+\alpha^2-4\,\beta\,\mu^{2}\,t^{2} \,,
\end{align}
where
\begin{align}\label{Coefficients}
\alpha&=\Delta^{2}_{0}+\left(\Delta^{2}_{1}+t^2\right)+\mu^2-{\cal{L}}^2 \,, \notag \\
\beta &=1-\left(1+\frac{\Delta^2_{0}}{t^{2}}\right)\frac{{\cal{L}}^2}{\mu^2} +\left(2+\frac{\Delta_{0}\,\Delta_{1}}{\mu\,t}\right)\frac{\Delta_{0}\,\Delta_{1}}{\mu\,t} \,.
\end{align}
Note that, in the expressions above, the explicit ${\bf \Gamma}_i$-dependence has been omitted for more clearness. Besides, it has been implied that ${\cal{L}}({\bf \Gamma}_i)=|{\cal{L}}_{\uparrow\downarrow}({\bf \Gamma}_i)|$. Note also that time-reversal symmetry in Eq.\,(\ref{Kinetic Hamiltonian}) requires that $t\left({\bf \Gamma}_{i}\right)=t^{*}\left({\bf \Gamma}_{i}\right)$, meaning that $t({\bf \Gamma}_{i})$ is a real number. This argument holds when $\mu\left({\bf \Gamma}_{i}\right)$ is real too, and so are the pairing energies $\Delta_{0}$ and $\Delta_{1}({\bf \Gamma}_i)$.

In order to induce topological band inversions, it is then crucial that the Bloch dispersion relation of the underlying crystal as described by Eq.\,(\ref{Kinetic Hamiltonian}) has an energy gap at the symmetry points ${\bf \Gamma}_{i}$, which implies $\epsilon_{\pm}({\bf \Gamma}_{i})=\pm |t({\bf \Gamma}_{i})|\neq0$. Otherwise, determinant (\ref{Determinant}) cannot change signs, since $t\left({\bf \Gamma}_{i}\right)=0$ leads to $\Det \tilde{H}({\bf \Gamma}_{i})=[V_{z}^{2} - \alpha ({\bf \Gamma}_{i})]^{2}\geq0$. Such a situation has been reported in the context of the Dirac cone merging transition in 2D materials such as graphene and few-layer black phosphorus [\onlinecite{dutreix2016laser},\,\onlinecite{dietl2008new,Montambaux:2009ff,kim2015observation}].

In the case of $s$-wave superconductivity ($\Delta_{1}=0$), determinant (\ref{Determinant}) becomes negative when $\beta \left({\bf \Gamma}_{i}\right)>0$, which equivalently reads
\begin{align}
\label{Critical Doping 1}
\mu>\mu_{c}=\sqrt{1+\left(\frac{\Delta_{0}}{t\left({\bf \Gamma}_{i}\right)}\right)^{2}}~ {\cal{L}}\left({\bf \Gamma}_{i}\right) \,.
\end{align}
As a result, it becomes mandatory to dope the system in the case of a diatomic-pattern crystal since ${\cal{L}}\left({\bf \Gamma}_{i}\right)\neq0$. This implies in particular that Majorana quasiparticles cannot occur at zero-energy in graphene, as already discussed from symmetry argument in Ref.\,[\onlinecite{2012PhST..146a4013C}]. The critical doping $\mu_{c}$ that is required to allow topological band inversions mainly depends on the strength of the Rashba spin-orbit interactions. The sign of determinant (\ref{Determinant}) finally turns out to be negative when $V^{2}_{-}<V^{2}_{z}<V^{2}_{+}$ where
\begin{align}\label{Critical Potential 1}
V_{\pm}^2=\Delta^{2}_{0}+t^{2}({\bf \Gamma}_{i})+\mu^2-{\cal{L}}^2({\bf \Gamma}_i)\pm2\,\mu \, t\left({\bf \Gamma}_{i}\right)\sqrt{1-\left(\frac{\mu_{c}}{\mu}\right)^{2}} \notag \,.
\end{align}
Because of the sublattice structure, the Rashba spin-orbit interaction is now involved in the topological band inversions at the symmetry points ${\bf \Gamma}_{i}$. This can be compared to what happens in monatomic-pattern crystals and noncentrosymmetric superconductors where $V^{2}_{\pm}=\Delta^{2}_{0}+[\epsilon({\bf \Gamma}_{i})\pm\mu]^{2}$ and where the Rashba spin-orbit only controls the magnitude of the bulk energy gap [\onlinecite{Sato:2010kb}].

When $\Delta_{0}=0$, similar conclusions hold. Indeed, $\beta \left({\bf \Gamma}_{i}\right)>0$ implies
\begin{align}
\mu>\mu_{c}= {\cal{L}}({\bf \Gamma}_{i}) \,,
\end{align}
so that the strength of the Rashba spin-orbit interactions fixes the minimal doping. The sign of determinant (\ref{Determinant}) is negative for Zeeman potentials that satisfy $V^{2}_{-}<V^{2}_{z}<V^{2}_{+}$ where
\begin{align}
V_{\pm}^2=\Delta^{2}_{1}({\bf \Gamma}_{i})+t^{2}({\bf \Gamma}_{i})+\mu^2-{\cal{L}}^2({\bf \Gamma}_i)\pm2\,\mu \, t\left({\bf \Gamma}_{i}\right)\sqrt{1-\left(\frac{\mu_{c}}{\mu}\right)^{2}} \notag \,.
\end{align}
One more time the Rashba spin orbit leads to a more restrictive condition because of the sublattice structure of the crystal.

\subsection{Topological phases with Majorana boundary quasiparticles}

\subsubsection{Dimerized Peierls chain}
Let us start with the case of the 1D dimerized Peierls crystal, as illustrated in Fig.\,\ref{Lattices}. This model was, for example, investigated by Su, Schrieffer, and Heeger, to explain the formation of topological solitons in polyacetylene, an organic semiconductor [\onlinecite{su1979solitons}]. Its electronic properties are described by Hamiltonian $\mathcal{H}_{0}$, as introduced in Eq.\,(\ref{Kinetic Hamiltonian}), for $t(k) = t\, (1+\alpha e^{-ik})$ and $\mu(k)=\mu$. Here $t$ and $\mu$ respectively denote the nearest-neighbor  hopping amplitude and the chemical potential. The dimensionless parameter $\alpha$ simulates the dimerization of the chain or, in other words, the existence of different intradimer and interdimer hopping amplitudes. Note moreover that momentum $k$ is assumed to be dimensionless and given in units of the lattice constant. The Rashba Hamiltonian $\mathcal{H}_{R}$ has been introduced in Eq.\,(\ref{Rashba Hamiltonian}) and here ${\cal{L}}_{\uparrow\downarrow}(k)=\lambda\,(1+\alpha e^{-ik})=-{\cal{L}}_{\downarrow\uparrow}^{*}(-k)$, where $\lambda$ controls the strength of the Rashba spin-orbit interactions. It is also assumed that the spin-singlet superconductivity is induced by proximity effect and $\Delta_{1}=0$.

\begin{figure}[t]
\centering
$\begin{array}{c}
\includegraphics[trim = 15mm 0mm 20mm 0mm, clip, width=6cm]{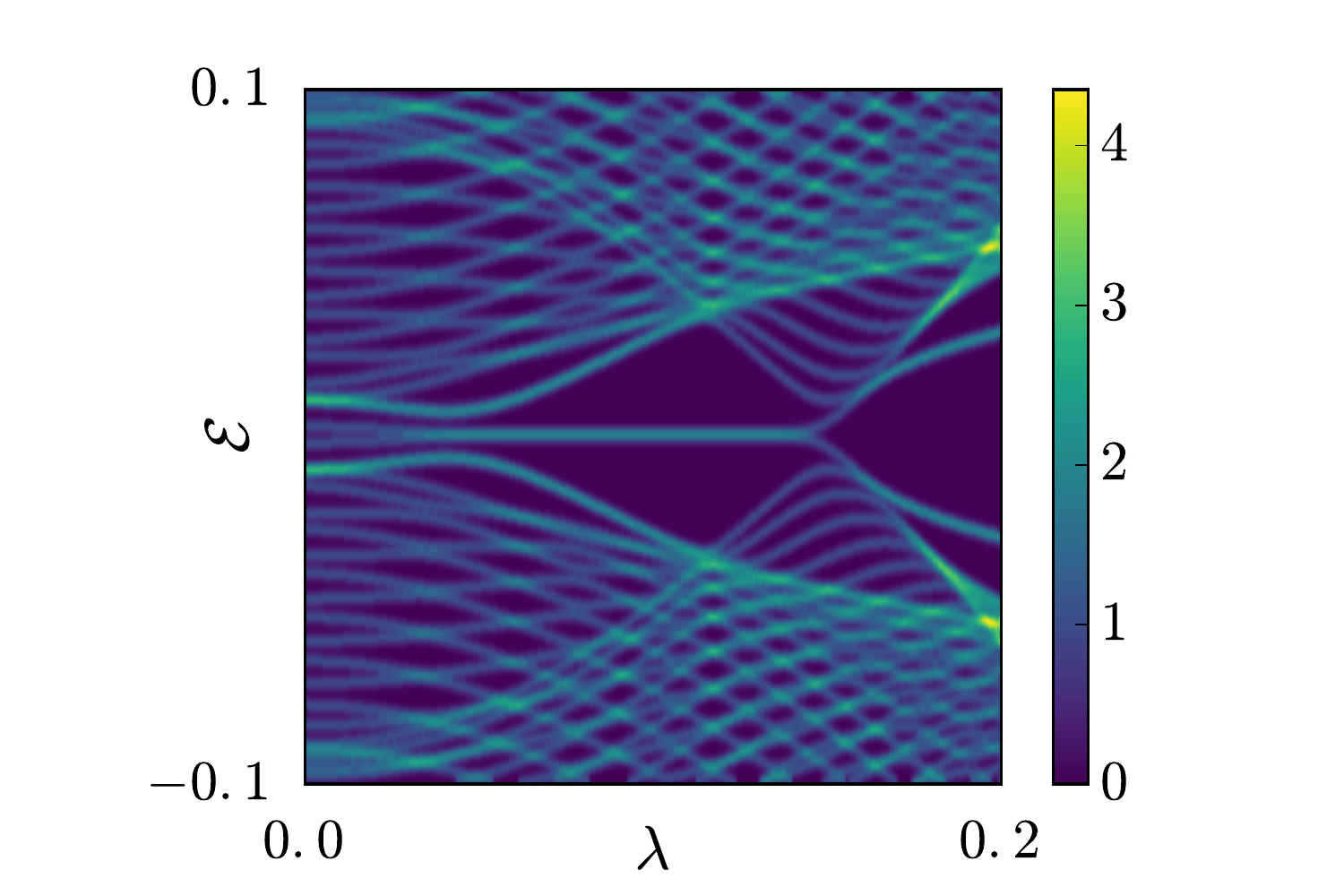}\\
\includegraphics[trim = 0mm 0mm 10mm 0mm, clip, width=5.5cm]{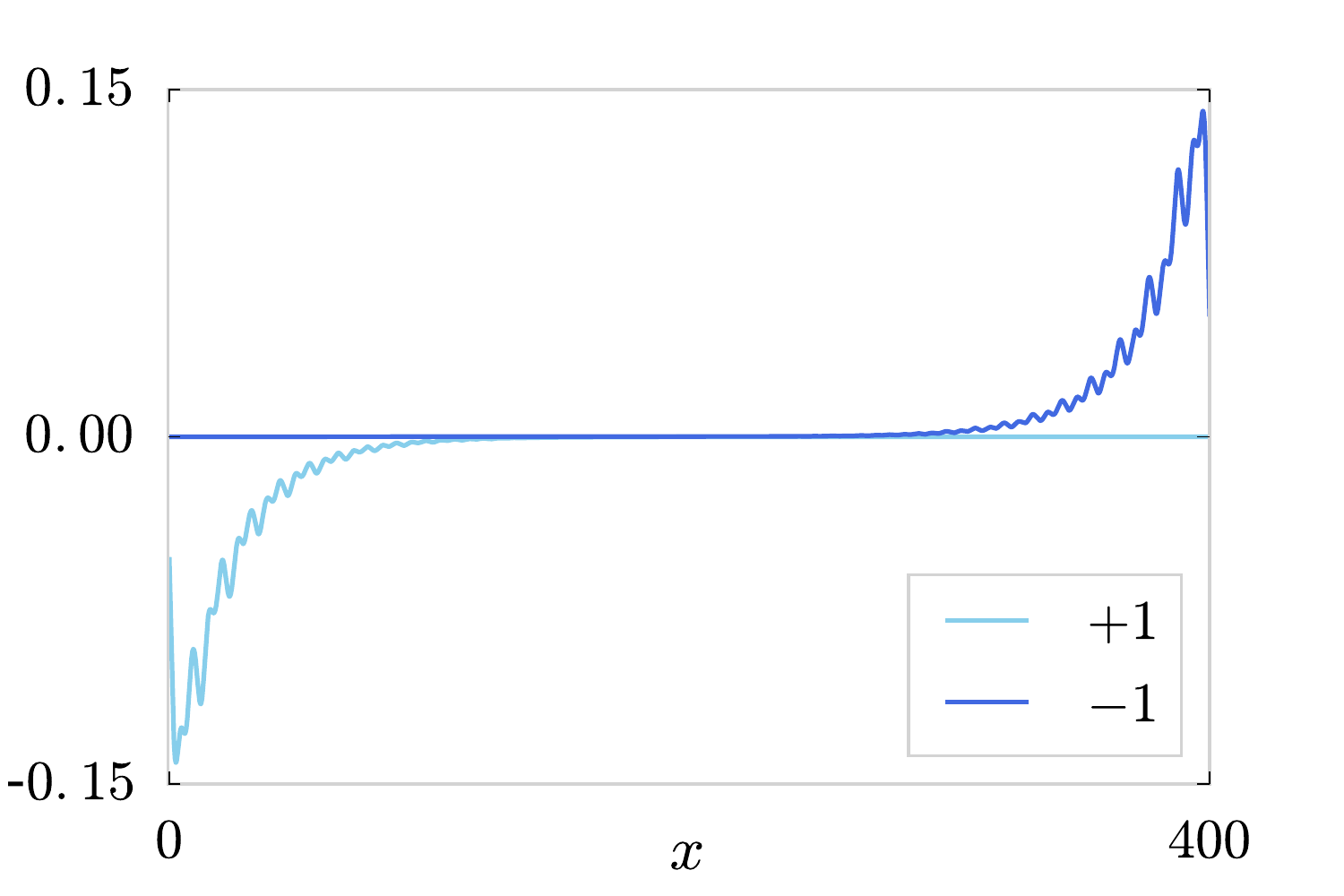}
\end{array}$
\caption{\small (Color online) Number of states as a function of energy $\mathcal{E}$ and spin-orbit strength $\lambda$ (top), and zero-energy Majorana polarization as a function of position $x$ (bottom) for a Peierls crystal ($\alpha=0.9$) made of $400$ sites. Parameters are such that $V_{z}=\mu=0.5$ and $\Delta_{0}=0.1$ for both plots and additionally $\lambda=0.1$ for the second one. Energy is given in units of the nearest-neighbor hopping amplitude $t$, while distance is given in units of the lattice constant.}
\label{1d DOS}
\end{figure}

When the doping satisfies condition (\ref{Critical Doping 1}), it becomes possible to induce topological band inversions, so that the parity product $\delta_{1}$ changes signs, similarly to the situation depicted in Fig.\,\ref{Parity Products}. The phase diagrams this leads to are shown in Fig.\,\ref{Zeeman Phase Diagram} and Fig.\,\ref{Supra Phase Diagram}. The $\mathbb{Z}_{2}$ topological invariant is exactly obtained from the analytical computation of the parity products $\delta_{0}$ and $\delta_{1}$. When $(-1)^{\nu}=\delta_{0}\delta_{1}=-1$, the 1D Bloch band structure is topologically nontrivial. It can be checked that the minimum doping required to reach the topological phases (dark purple areas) increases when increasing the strength of the Rashba spin orbit, in agreement with condition (\ref{Critical Doping 1}). More generally, those two figures clearly highlight that the topological properties of the band structure are directly affected the strength of the spin-flip process. This Rashba dependence in the case of a finite Peierls chain is also illustrated in Fig.\,\ref{1d DOS}. The two zero-energy states only exist for a certain range of the spin obit strengths. The figure also shows the Majorana polarization of the wave function, as defined in Refs.\,[\onlinecite{sticlet2012spin},\,\onlinecite{sedlmayr2015visualizing}], which confirms that the zero-energy modes are indeed Majorana boundary quasiparticles.

\subsubsection{(Stretched) Graphene}
The honeycomb lattice of graphene consists of a triangular Bravais lattice with a diatomic pattern, as depicted in Fig.\,\ref{Lattices}. The behavior of the $\pi$ electrons reveal a 2D semimetal and it can be described within a tight-binding approximation by the Hamiltonian of Eq.\,(\ref{Kinetic Hamiltonian}) with $t({\bf k}) = t\, (\alpha+ e^{-i{\bf k}\cdot{\bf a_{1}}}+e^{-i{\bf k}\cdot{\bf a_{2}}})$ and $\mu({\bf k})=\mu$. The nearest-neighbor hopping amplitude satisfies $t\simeq -3$\,eV [\onlinecite{reich2002tight}], and $\mu$ denotes the chemical potential. Momentum $k$ is assumed to be dimensionless and given in units of the lattice constant. Note that the topological prescription introduced above holds as long as the inversion symmetry between the two sublattices is not broken. Thus, the upcoming discussion could straightforwardly be generalized to distant neighbor hopping processes. The dimensionless parameter $\alpha$, which for instance simulates a uniaxial strain, controls the Dirac cone merging transition and the semi-relativistic phenomena it leads to in two dimensions [\onlinecite{dietl2008new},\,\onlinecite{Montambaux:2009ff},\,\onlinecite{dutreix2013friedel}]. Even though such a Lifshitz transition is unrealistic in graphene, contrary to black phosphorus [\onlinecite{kim2015observation}], reasonably stretching the graphene sheet turns out to be useful here, because it reduces the minimum doping required to obtain Majorana boundary quasiparticles. The Rashba Hamiltonian in Eq.\,(\ref{Rashba Hamiltonian}) is given by $\mathcal{H}_{R}=i\lambda\,\sum_{mn}({\bf d_{mn}}\times\boldsymbol{\sigma} \cdot {\bf e_{z}})_{\sigma\sigma'}a_{m\sigma}^{\dagger}b_{n\sigma'}+H.c.$, where vector ${\bf d_{mn}}$ connects two nearest-neighbor sites $m$ and $n$, $\boldsymbol{\sigma}$ is a vector whose components are the Pauli matrices, and ${\bf e_{z}}$ denotes a unit vector perpendicular to the honeycomb lattice. The Rashba spin-orbit constant has been estimated as $\lambda\simeq0.001--0.010$\,meV in graphene under electric field [\onlinecite{min2006intrinsic}\,,~\onlinecite{huertas2006spin}]. Spin-singlet superconductivity is assumed to be induced by proximity effect ($\Delta_{1}=0$), which may be achieved by growing a graphene layer on top of a superconducting thin film of Re(0001) [\onlinecite{tonnoir2013induced}]. Note that the prescription introduced above also holds when superconductivity arises from strong electron-electron interactions near the van Hove singularities ($\Delta_{1}\neq0$) [\onlinecite{PhysRevB.75.134512,PhysRevLett.98.146801,Black-Schaffer:2012os}].

\begin{figure}[t]
\centering
$\begin{array}{cc}
\includegraphics[trim = 0mm 0mm 10mm 0mm, clip, width=5.5cm]{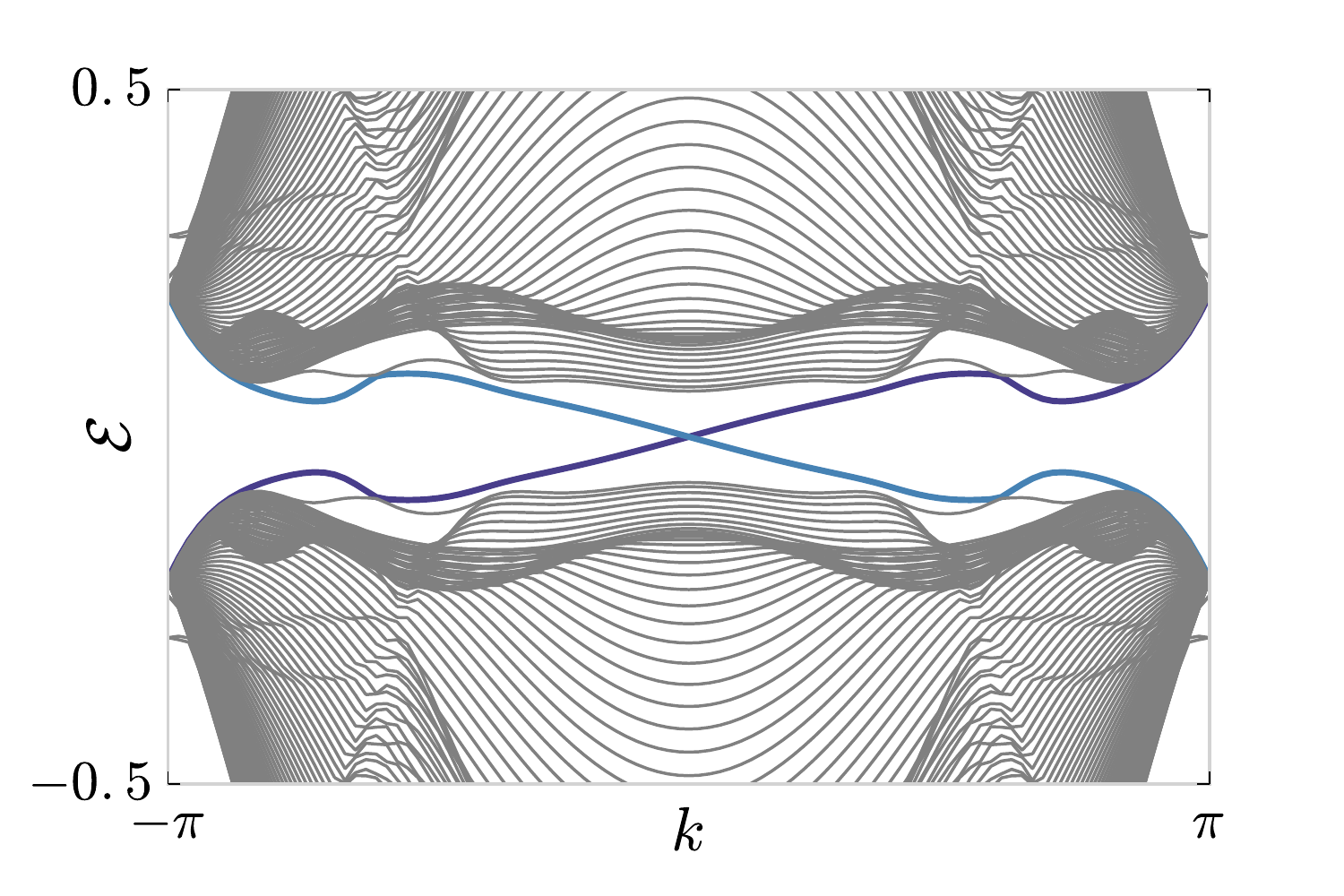}&
\includegraphics[trim = 64mm 0.mm 10mm 0mm, clip, width=3.03cm]{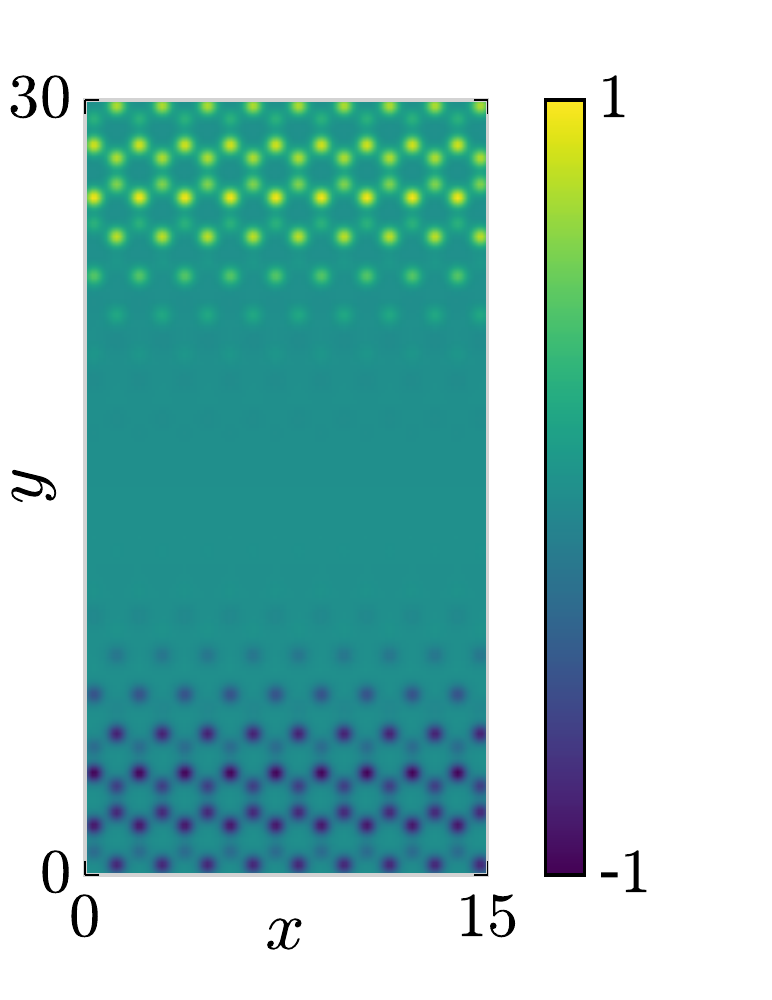}\\
\includegraphics[trim = 0mm 0mm 10mm 0mm, clip, width=5.5cm]{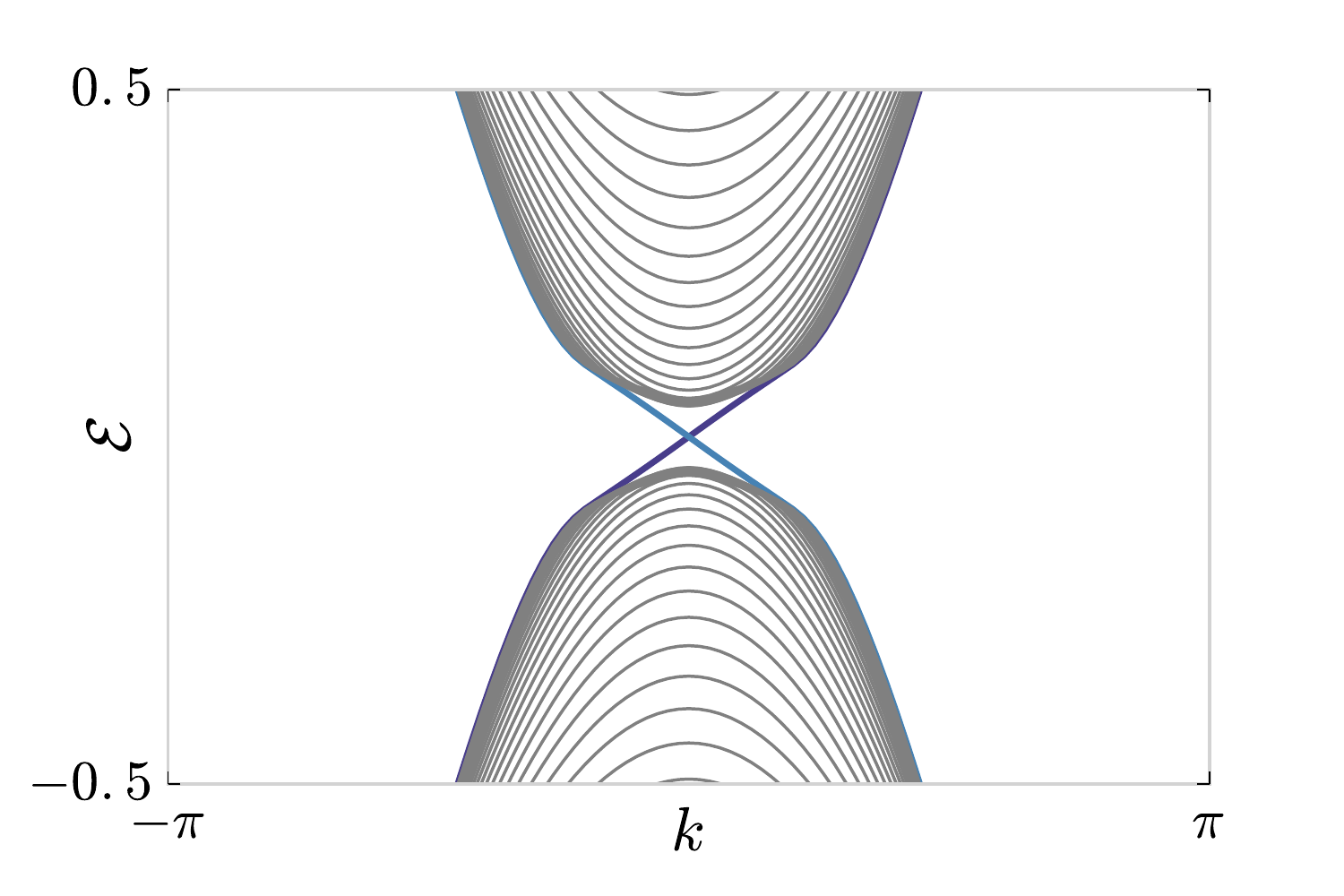}&
\includegraphics[trim = 64mm 0.mm 10mm 0mm, clip, width=3.03cm]{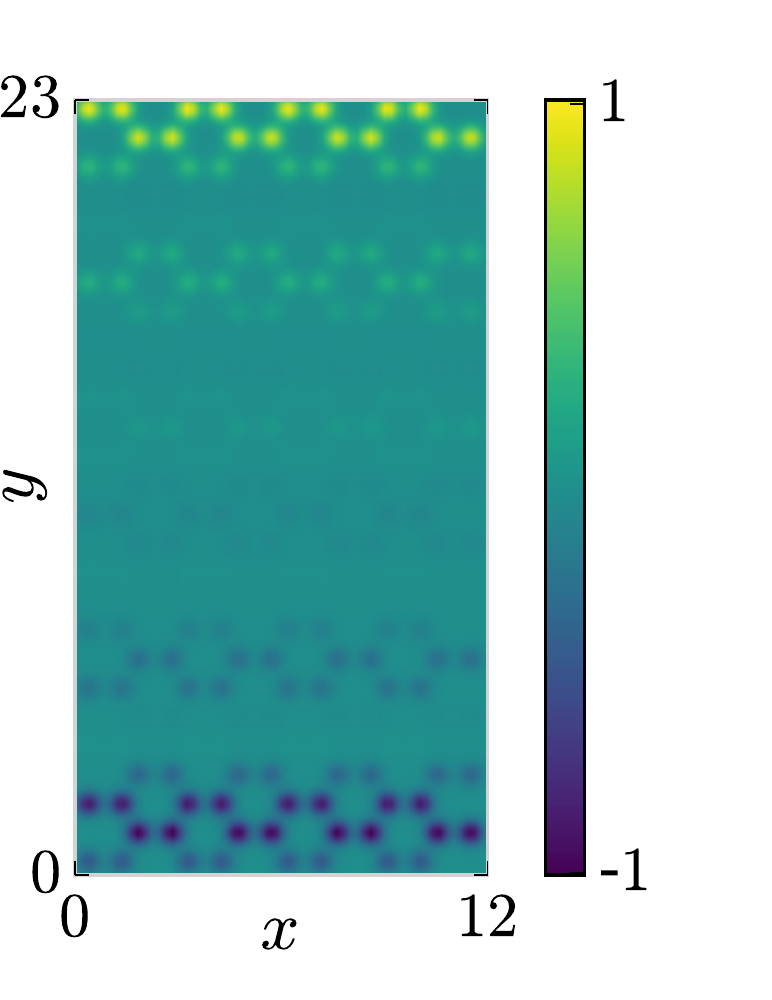}
\end{array}$
\caption{\small (Color online) Spectrum of a stretched graphene nanoribbon with zig-zag termination ($\alpha=1.3$, $\mu=0.7$) and polarization of the chiral Majorana boundary quasiparticles at zero-energy (first line, from left to right). Spectrum of a phosphorene nanoribbon with armchair termination ($\alpha=3.0$, $\mu=1.0$, $V_{E}=1.0$) and polarization of the chiral Majorana boundary quasiparticles at zero-energy (second line, from left to right). Other parameters correspond to $V_{z}=0.6$, $\Delta_{0}=0.5$, and $\lambda=0.1$. Energy is given in units of the nearest-neighbor hopping amplitude $t$ and distance in units of the lattice constant.}
\label{2d DOS}
\end{figure}

Topological band inversions may then be induced when the minimum doping condition given in Eq\,.(\ref{Critical Doping 1}) is fulfilled. In the case of isotropic graphene, such band inversions arise, for example, at the symmetry points ${\bf \Gamma_{1}}$, ${\bf \Gamma_{2}}$, and ${\bf \Gamma_{3}}$ all together, since for the three of them the van Hove singularities refer to the same energy. This corresponds to the situation illustrated in Fig.\,\ref{Parity Products}. The phase diagrams this situation leads to are depicted in the second lines of Fig.\,\ref{Zeeman Phase Diagram} and Fig.\,\ref{Supra Phase Diagram} for different strengths of the Rashba spin orbit. The Chern number $\nu$ necessarily corresponds to a non-trivial topological phase when $(-1)^{\nu}=\delta_{0}\delta_{1}\delta_{2}\delta_{3}=-1$. In particular, Fig.\,\ref{Supra Phase Diagram} shows that, in order to reach a topological phase, the Fermi level has to be fixed either in the vicinity of the van Hove singularity ($\mu\simeq t$), or in the vicinity of the top of the conduction band ($\mu\simeq 3t$), even though the latter turns out to be unrealistic. Importantly, these analytical phase diagrams confirm the numerical predictions discussed in Refs.\,[\onlinecite{Black-Schaffer:2012os},\,\onlinecite{dutreix:2014EPJB}]. Note that this feature of the doping near the van Hove singularities is also visible in Fig.\,\ref{Zeeman Phase Diagram}, when minimizing the Zeeman potential in the topological phase. Besides, it would be experimentally desirable to reduce the doping quite below the van Hove singularities. However, the minimum doping required to enable the system to enter the topological phase strongly depends on the strength on the Rashba spin-orbit, according to the phase diagrams and in agreement with condition (\ref{Critical Doping 1}). One may then reduce this critical doping by stretching the graphene sheet. Indeed, when increasing $\alpha$ up to $1.3$, this decreases the energy level at one of the symmetric points ${\bf \Gamma_{1}}$, ${\bf \Gamma_{2}}$, and ${\bf \Gamma_{3}}$, so that the three van Hove singularities are no longer equivalent [\onlinecite{dietl2008new},\,\onlinecite{Montambaux:2009ff},\,\onlinecite{dutreix2013friedel}]. As a consequence, if the Fermi level lies near the lowest van Hove singularities, i.e., $\mu=0.7t$ at ${\bf \Gamma_{3}}$ in our case, the system may still enter a topological phase. It is also possible to reduce this minimum doping by increasing the Zeeman potential, as long as this doping is larger than the critical one fixed by the spin-orbit strength in Eq.\,(\ref{Critical Doping 1}). Importantly when the Chern number satisfies $(-1)^{\nu}=-1$, there may exist zero-energy Majorana quasiparticles at the boundaries of the system. This is exemplified in Fig.\,\ref{2d DOS} with the Majorana polarization of the two zero-energy chiral modes that are located at the zig-zag edges of a stretched graphene nanoribbon.

\subsubsection{Phosphorene}
Phosphorene is a black phosphorus monolayer whose 2D structure consists of a puckered honeycomb lattice, i.e., a rectangular Bravais lattice with four non-equivalent atoms per unit cell [\onlinecite{liu2014phosphorene}]. The electronic properties of these material define an anisotropic semiconductor, which can be well explained in terms of a tight-binding model [\onlinecite{rudenko2014quasiparticle}]. Here we restrict our analysis to the nearest-neighbor approximation of this model, in which the anisotropy relies on two different hopping amplitudes that are $t\simeq-1.2$\,eV and $t'=-\alpha t$ with $\alpha\simeq 3$ [\onlinecite{rudenko2014quasiparticle}]. This description is detailed in Appendix\,\ref{Tight-binding model for phosphorene}. Importantly, because of the puckered structure of phosphorene, the non-equivalent electronic orbitals do not experience the same electric potential if an electric field is applied perpendicularly to the material. Two of the four orbitals experience an electric potential $V_{E}$, while the two others experience $-V_{E}$. For an electric potential $V_{E}\simeq1$\,eV, the two different Rashba spin-orbit hoppings have been estimated as $\lambda\simeq0.004$\,eV and $\lambda'=\alpha \lambda$ [\onlinecite{popovic2015electronic}]. As shown in Appendix\,\ref{Tight-binding model for phosphorene}, this description preserves inversion symmetry within the phosphorene sheet. Therefore, we can follow the same procedure as for the case of graphene, and look for the topological band inversions in the presence of spin-singlet superconductivity, which can be induced by electron doping [\onlinecite{shao2014electron},\,\onlinecite{Zhang:2017qq}].

The Fermi level is supposed to stay in the bottom of the conduction band centered around ${\bf \Gamma_{0}}$, which implies $\mu=t$. This condition looks like the one we have dealt with in graphene, but now $t$ is about three times smaller, so that we are considering a lower doping. Figures\,\ref{Zeeman Phase Diagram} and Fig.\,\ref{Supra Phase Diagram} depict the topological phases given by $(-1)^{\nu}=\delta_{0}\delta_{1}\delta_{2}\delta_{3}=-1$. Both figures shows that the effect of the Rashba spin-orbit is to increase the minimum doping necessary to reach the non-vanishing values of the Chern number $\nu$. The topological phase characterized by $(-1)^{\nu}=-1$ is associated to the existence of Majorana quasiparticles at the boundaries of a phosphorene nanoribbon. Figure\,\ref{2d DOS} shows the Majorana polarization of the chiral zero-energy modes at the edges of a nanoribbon with armchair terminations. Please note that the nearest-neighbor vectors have voluntarily been chosen as isotropic to depict the top view of phosphorene honeycomb lattice in Fig.\,\ref{2d DOS}.

\section{Conclusion}

This work has addressed the problem of existence of Majorana boundary quasiparticles in spin-singlet superconducting materials that have an underlying sublattice structure. It has especially focused on diatomic pattern crystals such as the dimerized Peierls chain, (stretched) graphene, and phosphorene, in the presence of spin-singlet superconductivity, time-reversal-symmetry breaking Zeeman splitting, and Rashba spin orbit. These systems, which fall into the BdG class D, have been shown to have an extra $\Pi$-symmetry associated to a unitary operator that anticommutes with the charge-conjugation operator. A general prescription has then been given in order to connect the parity of the energy bands at the $\Pi$-symmetry invariant momenta ${\bf \Gamma_{i}}$ to the topological invariant of the Bloch band structure. Because of the underlying sublattices, the Rashba spin-orbit does not vanish at the ${\bf \Gamma_{i}}$ points and may then be responsible for topological transitions, thus affecting the existence condition of Majorana boundary quasiparticles in the system. The topological phase diagrams this prescription leads to have finally been obtained analytically and exactly. They reveal that, in the cases of 1D and 2D semiconductors, the Majorana boundary quasiparticles are likely to emerge when the Fermi level lies in the vicinity of the bottom (top) of the conduction (valence) band. In the case of a semimetal such as stretched graphene, it turns out that the Majorana quasiparticles cannot result from band inversions at the ${\bf \Gamma_{i}}$ points for zero and low doping, that is, when the Fermi energy is associated to the Dirac points. It is nevertheless possible to obtain such boundary quasiparticles when the Fermi level lies in the vicinity of the van Hove singularities.

\begin{acknowledgments}
The author is very grateful to F. Pi\'echon for instructive discussions and would like to thank C. Bena and P. Delplace for useful comments.
\end{acknowledgments}

\bibliographystyle{apsrev4-1}
\bibliography{references}

\appendix
\newpage
\onecolumngrid

\section{Antisymmetric sewing matrix}
\label{Appendix Antisymmetric sewing matrix}
The transpose of the sewing matrix $\mathcal{B}$, as defined in the main text, is simply given by
\begin{align}
\mathcal{B}^{T}_{mn}({\bf k}) &= \mathcal{B}_{nm}({\bf k}) \notag \\
&=  \langle u_{n}(-{\bf k}) | \mathcal{P} \mathcal{C} | u_{m}^{*}(-{\bf k}) \rangle \notag \\
&=  \langle u_{m}(-{\bf k}) | \mathcal{C}^{\dagger}\mathcal{P}^{\dagger}| u_{n}^{*}(-{\bf k}) \rangle \notag \\
&=  \langle u_{m}(-{\bf k}) | \mathcal{C}\mathcal{P}| u_{n}^{*}(-{\bf k}) \rangle \notag \\
&=  -\langle u_{m}(-{\bf k}) | \mathcal{P}\mathcal{C}| u_{n}^{*}(-{\bf k}) \rangle \notag \\
&= - \mathcal{B}_{mn}({\bf k}) \,.
\end{align}
Therefore, the sewing matrix is antisymmetric for all values of ${\bf k}$.

\section{Berry connection}
\label{Appendix Berry connection}
The Berry connection defined over the $M$ energy bands fulfills
\begin{align}
\mathcal{A}\left({\bf k}\right) 
=& -i \sum_{n=1}^{M} \langle u_{n}({\bf k}) | \nabla_{\bf k} | u_{n}({\bf k}) \rangle \notag \\
=& -i \sum_{n=1}^{M} \langle u^{*}_{n}(-{\bf k}) |\mathcal{C}^{\dagger} \nabla_{\bf k} \mathcal{C} | u^{*}_{n}(-{\bf k}) \rangle \notag \\
=& -i \sum_{n=1}^{M} \langle u^{*}_{n}(-{\bf k}) | \mathcal{C}^{\dagger} \mathcal{P}^{\dagger} \nabla_{\bf k} \mathcal{P} \mathcal{C}| u^{*}_{n}(-{\bf k}) \rangle \notag \\
=& -i \sum_{m,n=1}^{M} \langle u^{*}_{n}(-{\bf k}) | \mathcal{C}^{\dagger} \mathcal{P}^{\dagger} | \nabla_{\bf k} |u_{m}(-{\bf k})\rangle \langle u_{m}(-{\bf k}) | \mathcal{P} \mathcal{C} | u^{*}_{n}(-{\bf k}) \rangle \notag \\
=& -i \sum_{m,n=1}^{M} \langle u_{m}(-{\bf k}) | \mathcal{P}\mathcal{C} | u^{*}_{n}(-{\bf k}) \rangle \langle u^{*}_{n}(-{\bf k})| \mathcal{C}^{\dagger} \mathcal{P}^{\dagger} | \nabla_{\bf k} |u_{m}(-{\bf k})\rangle \notag \\
&- i \sum_{m,n=1}^{M} \langle u^{*}_{n}(-{\bf k}) | \mathcal{C}^{\dagger} \mathcal{P}^{\dagger} |u_{m}(-{\bf k})\rangle  \nabla_{\bf k} \langle u_{m}(-{\bf k}) | \mathcal{P} \mathcal{C} | u^{*}_{n}(-{\bf k}) \rangle \notag \\
=& - \mathcal{A}(-{\bf k}) -i\Tr \left[ \mathcal{B}^{\dagger}({\bf k}) \nabla_{\bf k} \mathcal{B}({\bf k}) \right] \notag \\
=& - \mathcal{A}(-{\bf k}) -i \nabla_{\bf k} \ln\Det \mathcal{B}({\bf k}) \,.
\end{align}
This results in
\begin{align}
\label{Ln Determinant 1}
\mathcal{A}\left({\bf k}\right) + \mathcal{A}\left(-{\bf k}\right) &= -i \nabla_{\bf k} \ln \Det \mathcal{B}\left({\bf k}\right) ~.
\end{align}
Moreover, the Berry connections of negative- and positive-energy bands are related to each other in the following way:
\begin{align}
\mathcal{A}^{+}({\bf k}) &= i \sum_{n=N+1}^{2N} \langle u_{n}({\bf k}) | \nabla_{\bf k} | u_{n}({\bf k}) \rangle \notag \\
&= i \sum_{n=1}^{N} \langle u^{*}_{n}(-{\bf k}) | \mathcal{C}^{\dagger} \nabla_{\bf k} \mathcal{C} | u^{*}_{n}(-{\bf k}) \rangle \notag \\
&= i \sum_{n=1}^{N} \langle u^{*}_{n}(-{\bf k}) | \nabla_{\bf k} | u^{*}_{n}(-{\bf k}) \rangle \notag \\
&= i \sum_{n=1}^{N} \langle u_{n}(-{\bf k}) | \nabla_{-\bf k} | u_{n}(-{\bf k}) \rangle \notag \\
&= \mathcal{A}^{-}(-{\bf k})~, \notag
\end{align}
where ``$\pm$'' respectively refers to the positive- and negative-energy bands. This leads to $\mathcal{A}\left({\bf k}\right)=\mathcal{A}\left(-{\bf k}\right)$ and
\begin{align}
\mathcal{A}^{-}({\bf k}) + \mathcal{A}^{-}(-{\bf k}) = -\frac{i}{2} \nabla_{\bf k} \ln \Det \mathcal{B} ({\bf k}) \,,
\end{align}
when using relation (\ref{Ln Determinant 1}).

\section{$\mathbb{Z}_{2}$ topological invariant in 1d}
\label{Appendix topological invariant in 1d}
The $\mathbb{Z}_{2}$ topological invariant in one dimension is $e^{i\gamma_{BZ}}$, where the Zak phase satisfies
\begin{align}
\gamma_{BZ}&= \oint_{BZ} d{\bf k} \cdot \mathcal{A}^{-}\left({\bf k}\right) \notag \\
&= \int_{-{\bf \Gamma}_{1}}^{{\bf \Gamma}_{0}} d{\bf k} \cdot \mathcal{A}^{-}\left({\bf k}\right)
+ \int_{{\bf \Gamma}_{0}}^{{\bf \Gamma}_{1}} d{\bf k} \cdot \mathcal{A}^{-}\left({\bf k}\right) \notag \\
&= \int_{{\bf \Gamma}_{0}}^{{\bf \Gamma}_{1}} d{\bf k} \cdot \mathcal{A}^{-}\left(-{\bf k}\right)
+ \int_{{\bf \Gamma}_{0}}^{{\bf \Gamma}_{1}} d{\bf k} \cdot \mathcal{A}^{-}\left({\bf k}\right) \notag \\
&= \int_{{\bf \Gamma}_{0}}^{{\bf \Gamma}_{1}} d{\bf k} \cdot \mathcal{A}^{+}\left({\bf k}\right)
+ \int_{{\bf \Gamma}_{0}}^{{\bf \Gamma}_{1}} d{\bf k} \cdot \mathcal{A}^{-}\left({\bf k}\right) \notag \\
&= \int_{{\bf \Gamma}_{0}}^{{\bf \Gamma}_{1}} d{\bf k} \cdot \mathcal{A}\left({\bf k}\right) \notag \\
&= -\frac{i}{2} \int_{{\bf \Gamma}_{0}}^{{\bf \Gamma}_{1}} d{\bf k} \cdot \nabla_{\bf k} \ln \Det \mathcal{B} ({\bf k}) \notag \\
&= -i \ln \sqrt{\frac{\Det \mathcal{B}\left({\bf \Gamma_{1}}\right)}{\Det \mathcal{B}\left({\bf \Gamma_{0}}\right)}} \notag \\
&= -i \ln \sqrt{\Det \mathcal{B}\left({\bf \Gamma_{0}}\right) \, \Det \mathcal{B}\left({\bf \Gamma_{1}}\right)} \notag \\
&= -i \ln \left[ \Pfaf \mathcal{B}\left({\bf \Gamma_{0}}\right) \, \Pfaf \mathcal{B}\left({\bf \Gamma_{1}}\right) \right] \,.
\end{align}
This finally leads to
\begin{align}
e^{i\gamma_{BZ}} &= \Pfaf \mathcal{B}\left({\bf \Gamma_{0}}\right) \, \Pfaf \mathcal{B}\left({\bf \Gamma_{1}}\right) \,,
\end{align}
where $\Pfaf \mathcal{B}$ denotes the Pfaffian of the antisymmetric sewing matrix.

\section{$\mathbb{Z}$ topological invariant in 2d}
\label{Appendix topological invariant in 2d}
In two dimensions, the BdG symmetry class D is characterized by a $\mathbb{Z}$ topological invariant [\onlinecite{schnyder2008classification}], namely a first Chern number $\nu$. Its definition relies on the vanishing Berry curvature $\mathcal{F}({\bf k})=\nabla_{{\bf k}} \times \mathcal{A}({\bf k})=0$, which also satisfies $\mathcal{F}^{\pm}({\bf k})=\nabla_{{\bf k}} \times \mathcal{A}^{\pm}({\bf k})=\mathcal{F}^{\pm}(-{\bf k})$. Thus,
\begin{align}
\nu &= \frac{1}{2\pi} \int_{BZ} d^{2}k ~ \mathcal{F}^{-}\left({\bf k}\right) \notag \\
&= \frac{1}{\pi} \int_{S} d^{2}k ~ \mathcal{F}^{-}\left({\bf k}\right) \,,
\end{align}
where $S$ corresponds to half the two-dimensional BZ, as illustrated in Fig\,\ref{Lattices}.
Besides, the Berry phase along the oriented path $\mathscr{C}$ that encloses once surface $S$ is given by
\begin{align}
\gamma_{\mathscr{C}}&= \oint_{\mathscr{C}} d{\bf k} \cdot \mathcal{A}^{-}\left({\bf k}\right) \notag \\
&= \int_{\Gamma_{0}}^{\Gamma_{1}} d{\bf k} \cdot \mathcal{A}\left({\bf k}\right) + \int_{\Gamma_{2}}^{\Gamma_{3}} d{\bf k} \cdot \mathcal{A}\left({\bf k}\right) \notag \\
&= -i \ln \sqrt{\frac{\Det \mathcal{B} \left({\bf \Gamma_{1}}\right)}{\Det \mathcal{B} \left({\bf \Gamma_{0}}\right)}\frac{\Det \mathcal{B} \left({\bf \Gamma_{3}}\right)}{\Det \mathcal{B} \left({\bf \Gamma_{2}}\right)}} ~.
\end{align}

\section{Tight-binding description of phosphorene}
\label{Tight-binding model for phosphorene}
Because of the puckered honeycomb lattice of phosphorene, a nearest-neighbor description involves four electronic orbitals A$_{1}$, A$_{2}$, B$_{1}$, as introduced in Ref.\,[\onlinecite{dutreix2016laser}] for example. In the presence of perpendicular electric field and Zeeman splitting, the Bloch Hamiltonian matrix expressed in the basis $\{$ A$_{1\uparrow}$, B$_{1\uparrow}$, A$_{2\uparrow}$, B$_{2\uparrow}$, A$_{1\downarrow}$, B$_{1\downarrow}$, A$_{2\downarrow}$, B$_{2\downarrow}$ $\}$ is well approximated by
\begin{align}
H({\bf k})=
\left( \begin{array}{ll} 
K_{\uparrow\uparrow}({\bf k}) & L_{\uparrow\downarrow}({\bf k})\\
L_{\uparrow\downarrow}^{\dagger}({\bf k}) & K_{\downarrow\downarrow}({\bf k}) \\
\end{array} \right) \,,
\end{align}
where
\begin{align}
K_{\sigma\sigma}({\bf k})=
\left( \begin{array}{llll} 
\mu+V_{E}+\sigma V_{z} & 0 & \mathcal{F}_{1}({\bf k}) & \mathcal{F}_{2}({\bf k}) \\
0 & \mu-V_{E}+\sigma V_{z} & \mathcal{F}_{2}({\bf k}) & \mathcal{F}_{1}({\bf k})\,e^{i\varphi({\bf k})} \\
\mathcal{F}_{1}^{*}({\bf k}) & \mathcal{F}_{2}^{*}({\bf k}) & \mu+V_{E}+\sigma V_{z} & 0 \\
\mathcal{F}_{2}^{*}({\bf k}) & \mathcal{F}_{1}^{*}({\bf k})\,e^{-i\varphi({\bf k})} & 0 & \mu-V_{E}+\sigma V_{z}
\end{array} \right) \,,
\end{align}
and
\begin{align}
L_{\uparrow\downarrow}({\bf k})=
\left( \begin{array}{llll}
0 & 0 & {\cal{L}}_{1\uparrow \downarrow}({\bf k}) & {\cal{L}}_{2\downarrow \uparrow}^{*}({\bf k}) \\
0 & 0 & {\cal{L}}_{2\downarrow \uparrow}^{*}({\bf k}) & {\cal{L}}_{1\uparrow \downarrow}({\bf k})\,e^{i\varphi({\bf k})} \\
{\cal{L}}_{1\downarrow \uparrow}^{*}({\bf k}) & {\cal{L}}_{2\uparrow \downarrow}({\bf k}) & 0 & 0 \\
{\cal{L}}_{2\uparrow \downarrow}({\bf k}) & {\cal{L}}_{1\downarrow \uparrow}^{*}({\bf k})\,e^{-i\varphi({\bf k})} & 0 & 0
\end{array} \right) \,.
\end{align}
In the above matrices $\mu$ denotes the chemical potential, $V_{E}$ is the on-site potential shift induced by a perpendicular electric field, and $V_{z}$ refers to a Zeeman splitting potential. Besides, $\mathcal{F}_{1}({\bf k})=t (1+e^{ik_{x}a_{x}})$, $\mathcal{F}_{2}({\bf k})=\alpha\, t$, and $\varphi({\bf k})=-k_{x}a_{x}-k_{y}a_{y}$. The nearest-neighbor hopping has been estimated to be $t_{1}\simeq-1.2$\,eV, while $\alpha\simeq3$, $a_{x}\simeq3.4$\,$\AA$ and $a_{y}\simeq3.6$\,$\AA$ [\onlinecite{popovic2015electronic}]. By breaking inversion symmetry, the electric field is also responsible for Rashba spin-orbit that is described by
\begin{align}
{\cal{L}}_{1\uparrow \downarrow}({\bf k})&=-i\lambda \left[ (d_{1y}+id_{1x}) + (d_{1y}-id_{1x})e^{ik_{x}a_{x}} \right] \\ \notag
{\cal{L}}_{1\downarrow \uparrow}({\bf k})&=-{\cal{L}}_{1\uparrow \downarrow}^{*}(-{\bf k}) \notag \\
{\cal{L}}_{2\uparrow \downarrow}({\bf k})&=-i\alpha\lambda d_{2y} \notag \\
{\cal{L}}_{2\downarrow \uparrow}({\bf k})&=-{\cal{L}}_{2\uparrow \downarrow}^{*}(-{\bf k}) \,,
\end{align}
where $d_{1x}$, $d_{1y}$, and $d_{2y}$ are components of the nearest-neighbor vectors and $\lambda\simeq0.004$\,eV for $V_{E}\simeq1.3$\,eV, accordingly to Ref.\,[\onlinecite{popovic2015electronic}]. Crucially, the spinless description of phosphorene characterized by diagonal blocks of the type $K_{\sigma\sigma}({\bf k})$ is still invariant under inversion symmetry since
\begin{align}
\mathcal{I} K_{\sigma\sigma}({\bf k}) \mathcal{I} = K_{\sigma\sigma}(-{\bf k})
\end{align}
where the inversion operator reads
\begin{align}
\mathcal{I}=
\left( \begin{array}{llll}
0 & 0 & 1 & 0 \\
0 & 0 & 0 & 1 \\
1 & 0 & 0 & 0 \\
0 & 1 & 0 & 0
\end{array} \right) \,.
\end{align}
Besides $\mathcal{I} \, L_{\sigma \sigma'}({\bf k}) \, \mathcal{I} = -L_{\sigma \sigma'}(-{\bf k})$. Consequently, one can straightforwardly follow the prescription introduced for diatomic pattern sublattices in Sec.\,\ref{Application to diatomic-pattern crystals}, and apprehend the band structure topology from band inversions.

\end{document}